\documentclass[preprintnumbers,prd,onecolumn,floatfix,superscriptaddress,nofootinbib]{revtex4}
\usepackage{graphicx}
\usepackage{epsfig}
\usepackage{bm}
\usepackage{amssymb}
\usepackage{float}
\usepackage{amsmath}
\usepackage{subfigure}
\usepackage{dcolumn}
\usepackage{cancel}
\usepackage[colorlinks]{hyperref}
\usepackage[usenames,dvipsnames]{color}

\hypersetup{
breaklinks=true,pdfstartview={FitH}, colorlinks=true, linkcolor=blue, citecolor=red, filecolor=magenta,    
urlcolor=blue, anchorcolor=green, linktocpage=true}

\def\doi{http://doi.org}

\begin{document}

\title{Analysis with observational constraints in $ \Lambda $-cosmology in $f(R,T)$ gravity}
\author{Ritika Nagpal}
\email{ritikanagpal.math@gmail.com}
\affiliation{Department of Mathematics, Netaji Subhas Institute of Technology, Faculty of Technology, University of Delhi, New Delhi-110078, India}
\author{S. K. J. Pacif}
\email{shibesh.math@gmail.com}
\affiliation{Centre for Theoretical Physics, Jamia Millia Islamia, New Delhi 110025, India}
\author{J. K. Singh}
\email{jainendrrakumar@rediffmail.com}
\affiliation{Department of Mathematics, Netaji Subhas Institute of Technology, Faculty of Technology, University of Delhi, New Delhi-110 078, India}
\author{Kazuharu Bamba}
\email{bamba@sss.fukushima-u.ac.jp}
\affiliation{Division of Human Support System, Faculty of Symbiotic Systems Science, Fukushima University, Fukushima 960-1296, Japan}
\author{A. Beesham}
\email{beeshama@unizulu.ac.za}
\affiliation{Department of Mathematical Sciences, University of Zululand, Kwa-Dlangezwa 3886, South Africa}

\begin{abstract}
An exact cosmological solution of Einstein field equations (EFEs) is derived for a dynamical vacuum energy  in $f(R,T)$ gravity for Friedmann-Lemaitre-Robertson-Walker (FLRW) space-time. A parametrization of the Hubble parameter is used to find a deterministic solution of EFE. The cosmological dynamics of our model is discussed in detail. We have analyzed the time evolution of physical parameters and obtained their bounds analytically. Moreover, the behavior of these parameters are shown graphically in terms of redshift $`z'$. Our model is consistent with the formation of structure in the Universe. The role of the $f(R,T)$ coupling constant $\lambda$ is discussed in the evolution of the equation of state parameter. The statefinder and Om diagnostic analysis is used to distinguish our model with other dark energy models. The maximum likelihood analysis has been reviewed to obtain the constraints on the Hubble parameter $H_0$ and the model parameter $n$ by taking into account the observational Hubble data set $H(z)$, the Union 2.1 compilation data set $SNeIa$, the Baryon Acoustic Oscillation data $BAO$, and the joint data set $H(z)$ + $ SNeIa$ and $H(z)$ + $SNeIa$ + $BAO $. It is demonstrated that the model is in good agreement with various observations.\\
\end{abstract}

\maketitle

\section{Introduction}

\qquad The late time elusive behavior of the Universe is one of the major challenges in modern cosmology. The current observational data of $ SNeIa $ confirms the late time cosmic speed up of the Universe. At present, much observational data is in support of the current acceleration of the Universe \cite{per, rie, perc, ste}. In order to explain this faster rate, a new form of energy is needed in the universe which has some anti-gravitational effect that drives the acceleration. This distinct type of energy with negative pressure is termed dark energy (DE) \cite{spe, ade, eis}. According to the Planck mission team, it is estimated that the Universe is composed of three main components, $ 4.9\% $ ordinary matter, $ 26.8\% $ dark matter (DM) and $ 68.3\% $ DE. The DM and DE are really different in nature. DM is attractive and responsible for formation of structure and clustering of the galaxies, whereas DE seems to be some kind of energy intrinsic to empty space which keeps getting stronger with time. There are multiple ideas on DE: one idea is that DE is a property of space itself, or some kind of dynamic energy fluid which has some opposite effects on the Universe to ordinary energy and matter. Although DE is a popular explanation for the expansion mystery supported by many observational experiments, there remain many unanswered questions.\\

The two main models proposed in literature to explain the nature of DE are cosmological constant $ \Lambda $ \textit{i.e.} assuming a constant energy density filled in space homogeneously, and scalar field model, which considers a dynamical variable energy density in space-time. The simplest and most favorable candidate of DE is the Einstein cosmological constant $ \Lambda $ \cite{var, peb} which works as a force that counteracts the force of gravity. Adding the cosmological constant $ \Lambda $ to EFE of the FLRW metric leads to $\Lambda CDM$ model which serves as the agent of an accelerating Universe. In spite of its theoretical and  phenomenological problems \cite{wei}, the $ \Lambda CDM $ model has been referred to as the most efficient answer to the question of cosmic acceleration in many aspects because of its consistency with observations.  According to the GR, the equation of state (EoS) in cosmology specifies the expansion rate of the Universe. Nowadays, the great attempt in observational cosmology is to analyse the EoS $\omega =\frac{p}{\rho }$ of various DE models, where $ \rho $ and $ p $ are the energy density and isotropic pressure of the fluid. The quintessence and phantom models which are dynamic scalar fields, are two specific cases of dark energy models having EoS parameter $ \omega >-1 $ and $ \omega <-1 $ respectively. The first scenario of quintessence model was proposed by Ratra and Peebles \cite{rat}. The Quintessence model differs from $ \Lambda CDM $ in explanation of DE as quintessence model is dynamic that changes with respect to time unlike $ \Lambda $ which always stays constant \cite{sam, cald}. The phantom model \cite{phant1, phant2, phant3} could cause a big rip in the Universe due to the growing energy density of DE \cite{par,noj,ast}. Also number of other scalar fields DE models have been proposed as spintessence \cite{boy}, k-essence \cite{arm, chiba}, quintom \cite{feng}, tachyon \cite{sen-tach,pad}, Chameleon \cite{kho} having EoS parameter $ \omega \in (-1,0) $ \textit{etc}. Another class of alternative idea to come up with the theory of dark fluid that unifies both DM and DE as a single phenomenon \cite{chap}. In order to understand the behavior of Chaplygin gas in detail, one can review some excellent work published by Singh \textit{et al.}  \cite{sin2,sin5,sin6}. In addition ,holographic dark energy (HDE) is also a suitable choice for DE (among other alternatives) that might be originated from the quantum fluctuations of space-time \cite{cop,myr,sami2,yoo}. \\

Another possibility of DE that might affect the expansion history of the Universe concerns the dissipative phenomena
in the form of bulk and shear viscosity. After the discovery of acceleration of the Universe, the concept of viscous
cosmology has been reconsidered again. Some recent papers provide an application of viscous cosmology to the accelerating
Universe \cite{ber1, ben}. For an isotropic and homogeneous cosmic expansion, the negative pressure of a bulk viscous
fluid might play the role of an exotic fluid and could account for the effects usually attributed to DE. It is observed that the primordial inflation and present cosmic accelerated expansion can be achieved in the Universe by the inclusion of viscosity concepts in which various cosmic aspects have been considered \cite{ber}. There are a number of articles in the literature which discuss the DE phenomenon as an effect of dissipative processes such as bulk viscosity, which has been
thoroughly studied in a cosmic medium (for a detailed review see \cite{ber2,ber3}).\\

In the other direction, the accelerating expansion of the Universe can be revealed by modifying the Einstein–Hilbert
action. The standard Einstein Lagrangian can be modified by replacing the scalar curvature R with some arbitrary function
of R; this is known as $ f(R) $ gravity. Moreover, the replacement of Ricci scalar $R$ with scalar torsion $T$ is known as $ f(T) $ gravity and with Gravitational constant $G$ is known as $ f(G) $ gravity. Many other modifications of underlying geometry can cause a different modified theory to GR. Among the wide range of alternative ideas of modified gravity, $ f(R) $ gravity theory is the most viable alternative theory of gravity \cite{tho}. $ f(R) $ gravity is considered good on large scales, but fails to show consistency on some of the observational tests, \textit{e.g.} on rotation of the curved spiral galaxies \cite{chi,olmo}, on solar system regime \cite{capo,eri}. A more generic extension of $ f(R) $ gravity could be considered, as $ f(R,Sm) $, where the matter Lagrangian $ S_{m} $ is a function of trace $ T $ of energy momentum tensor and is taken as $ f(R,T) $ gravity \cite{har}. The main reason to introduce the term $ T $ is to take the quantum effects and exotic imperfect fluids in to account, and $ f(R,T) $ gravity is also capable of explaining the late time cosmic speed up. Some observational tests \cite{mry1,mor1} have been applied to $ f(R,T)$ gravity in order to resolve the issues entailed by $ f(R) $ gravity. To understand $ f(R,T) $ theory in detail, one may refer to some excellent work \cite{Alvarenga:2013syu,Sharif:2012zzd,Houndjo:2011tu,Momeni:2011am,Yousaf:2016lls, mor2, alve, sin1, sin3, you, das, sin4}. For recent reviews on modified gravity theories, see, for instance, \cite{Nojiri:2010wj, Capozziello:2011et, Nojiri:2017ncd, Capozziello:2010zz, Bamba:2015uma}\\

Shabani \textit{et al.} \cite{shab1} have studied minimal $ g(R) + h(T ) $ Lagrangian, a pure non–minimal $ g(R)h(T) $ Lagrangian and non–minimal $ g(R) (1 + h(T)) $ Lagrangian in $ f(R,T)$ modified gravity with a dynamical systems approach against the background of FLRW metric. Shabani \textit{et al.}  have discussed the cosmological and solar system consequences, non-interacting generalized Chaplygin gas (GCG) with the baryonic matter, late time solutions of $ \Lambda CDM $ subclass of $ f(R,T) $  gravity using dynamical system of approach, late time cosmological evolution of the Universe in $ f(R,T) $  gravity with minimal curvature-matter coupling via considering linear perturbations in the neighborhood of equilibrium, and bouncing cosmological models against the background of $ f(R,T)=R + h(T) $ gravity in FLRW metric with a perfect fluid \cite{shab2, shab3, shab4, shab5, shab6, shab7}. Singh et al. \cite{sin7} have studied a bouncing Universe in the framework of $ f(R,T) $ gravity using a specific form of the Hubble parameter.\\

In the Palatini formalism of $ f(R) $ gravity called $ \Lambda (T) $ gravity, first proposed by Poplawski \cite{pop}, one considered as the most general case where a $ \Lambda $-term is present in the general gravitational Lagrangian, which is taken as a function of $ T $, where $ T $ being the trace of energy momentum tensor. Moreover, the Palatini $ f(R) $ gravity can be brought back when we ignore the pressure dependent term from $ \Lambda (T) $ gravity. Also, the dynamical cosmological constant $ \Lambda $ is supported by this theory to solve the cosmological constant problem \cite{wein} and it is in good agreement with $ \Lambda (T) $ gravity. A detailed review of $ \Lambda (T) $ cosmology in $ f(R,T) $ modified gravity can be found in \cite{sahoo, prad, nasr}. Bamba \textit{et al.} \cite{kbam} have studied various types of dark energy models \textit{e.g.} $ \Lambda CDM $, pseudo-rip Universes, little rip Universes, quintessence and phantom cosmological models with Type-I, II, III, IV, and non-singular DE models.\\  

In this paper, the work is organized as follows: Sect. 1 provides a brief introduction on dark energy and alternative
ideas to cosmic acceleration. In Sect. 2, we review the derivation of the field equations with variable cosmological
parameter and obtain exact solutions to the EFE using a specific parametrization of the Hubble parameter. In Sect. 3, we
discuss the dynamics of the obtained model and briefly analyze the behavior of the geometrical and physical parameters
with the help of some graphical representations. In Sect. 4, we study the energy conditions and perform the diagnostic
analysis for our model, and in Sect. 5, we observe the consistency of our model with some cosmological observations.
Finally, we conclude with our results in Sect. 6.

\section{Basic equations and its solutions}

\subsection{Field equations in $f(R,T)$ gravity}

\qquad $ f(R,T) $ gravity \cite{har} is a more generic extended theory of $ f(R) $ gravity or more precisely GR which explains the curvature-matter coupling in the Universe. The formalism of $ f(R,T)$ model depends on a term which is generally considered as a source in the Lagrangian matter $ S_m $. The action of $ f(R,T) $ gravity is defined as

\begin{equation}  \label{1}
S = \int\Big(\frac{1}{16\pi G} f(R,T)+S_m\Big)\sqrt{-g} dx^4.
\end{equation}
In the above action, we consider the functional form of $ f(R,T)=f_1(R)+f_2(T) $, sum of two independent functions of Ricci scalar and trace of energy momentum tensor respectively. We assume the forms of $
f_1(R)=\lambda R $ and $ f_2(T)=\lambda T $, where $ \lambda $ is any arbitrary coupling constant of $ f(R,T) $ gravity.\\

On taking variation of (\ref{1}) with respect to $g_{ij}$ and neglecting the boundary terms, we have 
\begin{equation}  \label{2}
f^{\prime }_1(R)R_{ij}-\frac{1}{2}(f_1(R)+f_2(T))g_{ij}+(g_{ij}\Box-\nabla_i
\nabla_j) f^{\prime }_1(R)=8\pi T_{ij}-f^{\prime }_2(T)T_{ij}-f^{\prime
}_2(T)\theta_{ij},
\end{equation}
where the prime denotes the derivative with respect to the argument, and the operator $\Box$ defined above is De Alembert's operator ($\Box\equiv \nabla^i\nabla_i$). The term $\theta_{ij}$ is defined as 
\begin{equation}\label{theta}
\theta_{ij}\equiv g^{lm} \frac{\delta T_{lm}}{\delta g^{ij}}.
\end{equation}
Also if the matter content filling in the Universe shows a perfect fluid behavior then in this case $\theta_{ij}$ becomes $\theta_{ij}=-2T_{ij}-p g_{ij}$, matter Lagrangian density $ S_m $ can be considered as $S_m=-p$, and the energy momentum tensor (EMT) takes the form $T_{ij}=(\rho+p)u_i u_j-p g_{ij}$. Here, $u^i=(0,0,0,1)$ is the $4$-velocity vector which satisfies the condition $u^iu_i=1$ and, $u^i\nabla_j u_i=0$ in a co-moving coordinate system. \\

Using the values of functions $f(R)$ and $f(T)$ in Eq.(\ref{2}), where $(g_{ij}\Box -\nabla _{i}\nabla _{j})\lambda =0$, the field Eq.(\ref{2}) takes the form 
\begin{equation}
G_{ij}=R_{ij}-\frac{1}{2}Rg_{ij}=\left( \frac{8\pi +\lambda }{\lambda }%
\right) T_{ij}+(p+\frac{1}{2}T)g_{ij}\text{.}  \label{3}
\end{equation}

The Einstein field equations with cosmological constant (in units of $G=c=1$) in the general theory of relativity is
\begin{equation}
G_{ij}=8\pi T_{ij}+\Lambda g_{ij},  \label{4}
\end{equation}%
and comparing Eqs.(\ref{3}) and (\ref{4}) by taking a non-negative small value of the arbitrary coupling constant $\lambda $ such that the signs of RHS of Eqs.(\ref{3}) and (\ref{4}) remain the same. Thus
we have 
\begin{equation}
\Lambda \equiv \Lambda (T)=(p+\frac{1}{2}T),  \label{5}
\end{equation}%
which regards the effective cosmological constant $\Lambda$ as a function of the trace $ T $ \cite{pop}. Therefore, the EMT yields
\begin{equation}
\Lambda =\Lambda (T)=\frac{1}{2}(\rho-p).  \label{6}
\end{equation}%
\qquad

Consider the flat FLRW metric against the background which expresses a curvature-less homogeneous and isotropic Universe as
\begin{equation}
ds^{2}=dt^{2}-a^{2}(t)(dx^{2}+dy^{2}+dz^{2}),  \label{7}
\end{equation}%
where $a(t)$ is the expansion scale factor.\newline

In the background of the metric (\ref{7}) in the $ f(R,T) $ gravity for $ \Lambda (T)$ cosmology, EFEs (\ref{3}) yield the following two independent equations: 
\begin{equation}
3H^{2}=\left( A+\frac{1}{2}\right) \rho -\frac{1}{2}p,  \label{8}
\end{equation}%
\begin{equation}
3H^2+2\dot{H}=-\left( A+\frac{1}{2}\right)p+\frac{1}{2} \rho, \label{9}
\end{equation}%
where $ A=\frac{8\pi +\lambda }{\lambda } $, $ H=\frac{\dot{a}}{a} $ is the Hubble parameter which measure the fractional rate of change of scale factor $ a(t) $ and an overhead dot indicates the time derivative. In the next section, we solve the cosmological equations with a particular parametrization of Hubble parameter.

\subsection{Parametrization of $ H $ and exact solution}

\qquad The composition of the above two evolution Eqs. (\ref{8}) and (\ref{9}) involves three unknowns $ a $, $ \rho $ and $ p $. In order to accomplish a unique and consistent solution of the field equations, an additional constrain equation is needed. In general, the EoS parameter for the matter content of the Universe is considered as a supplementary condition. But there are other approaches too, which have been discussed by many authors regarding the \textit{parametrization} of the cosmological variables involved in the field equations, \textit{e.g.} the Hubble parameter, deceleration parameter, EoS parameter, energy density, pressure, and the cosmological constant \cite{lin,skjp}. From Eqs. (\ref{8}) and (\ref{9}) $ \rho $, $ p $ and $ \omega $  can also be represented in terms of $ H $ and $ q $ as

\begin{equation}
\rho =\frac{1}{(A+1)}\left[3+\frac{(q+1)}{A}\right] H^{2},  \label{z1}
\end{equation}

\begin{equation}
p=\frac{1}{(A+1)}\left[ -3+\frac{(q+1)(2A+1)}{A}\right] H^{2},  \label{z2}
\end{equation}

\begin{equation}
\omega=\frac{(2A+1)q-(A-1)}{(3A+q+1)}.  \label{omega}
\end{equation}

Here, $q$ is a dimensionless quantity which is a measure of cosmic acceleration in the Universe and is called deceleration parameter (DP). $q<0$ indicates the accelerated expansion in the Universe, whereas $q>0$ shows the expansion in the Universe as it is decelerated. The DP in terms of the scale factor $a$ and the Hubble parameter $H$ is defined as 
\begin{equation}\label{dp}
q=-\frac{\ddot{a}a}{\dot{a}^{2}}=-1-\frac{\dot{H}}{H^{2}}.
\end{equation}

From Eqs. (\ref{z1}) and (\ref{z2}), we obtain the solution for $q$ or $H$ explicitly. Eq. (\ref{omega}) represents the general expression of the EoS in the presence of $f(R,T)$ gravity. As recent astronomical observations acknowledge, the accelerating phase of the Universe was preceded by a decelerating phase. Taking the phase transition scenario in our present study, we choose an appropriate parametrization of the Hubble parameter $ H $ \cite{jps, ban}:  
\begin{equation}
H(a)=\alpha (1+a^{-n}),  \label{9a}
\end{equation}%
where $ \alpha >0 $ and $ n>1 $ are constants, called as model parameters, which are to be constrained through observations. Integrating Eq. (\ref{9a}), we obtain the scale factor in explicit form as

\begin{equation}  \label{9b}
a(t)=(e^{n \alpha t}-1)^{\frac{1}{n}}+c,
\end{equation}
where, we get the point type singularity at $ t=0 $ by taking arbitrary integration constant $ c $ as zero. The deceleration parameter $ q $ is given by

\begin{equation} \label{10d}
q(t)=\frac{n}{e^{n\alpha t}}-1.  
\end{equation}%
Using Eqs. (\ref{9a}) and (\ref{10d}) in Eqs. (\ref{z1}), (\ref {z2}) and (\ref {omega}), we obtain the physical parameters as 
\begin{equation} \label{10e}
\rho (t)=\frac{\lambda}{(8\pi+2\lambda)}\left[3+\frac{n\lambda}{e^{n\alpha t}(8\pi +\lambda)}\right]\frac{\alpha^2 e^{2n\alpha t}}{(e^{n\alpha t}-1)^2},  
\end{equation}

\begin{equation} \label{10p}
p(t)=\frac{\lambda}{(8\pi+2\lambda)}\left[-3+\frac{n(16+3\lambda)}{e^{n\alpha t}(8\pi +\lambda)}\right]\frac{\alpha^2 e^{2n\alpha t}}{(e^{n\alpha t}-1)^2},
\end{equation}

\begin{equation}\label{10g}
\omega (t)=\frac{-3(8\pi+\lambda)e^{n \alpha t}+n(16\pi+3\lambda)}{3(8\pi+\lambda)e^{n \alpha t}+n \lambda},  
\end{equation}
and 
\begin{equation}\label{10h}
\Lambda (t)=\left[\frac{3\lambda}{(8\pi+2\lambda)}-\frac{n\lambda}{e^{n\alpha t}}\right]\frac{\alpha^2 e^{2n\alpha t}}{(e^{n\alpha t}-1)^2}.
\end{equation}

\subsection{Bounds on the cosmological parameters}

\qquad Here, we evaluate the cosmological parameters at two extreme values of time $ t\rightarrow 0 $ and $ t\rightarrow \infty$, to examine the behavior of the model at the initial singularity as well as late time (see Table I). From Table 1, we can have a range of these cosmological parameters which depend on the parameter $ n $ and the $ f(R,T) $ coupling constant $ \lambda $, where $ \lambda=\frac{8\pi}{A-1} $. By choosing suitable values of $ n $ and $ \lambda $, we can explain the history of the expansion in terms of the various cosmological parameters. The role of the $ f(R,T) $ coupling constant $ \lambda $ can be seen clearly from Table 1. The Universe starts with infinite velocity and a finite acceleration and expands indefinitely with constant velocity and constant acceleration. The energy density and isotropic pressure start from infinitely large values at the time of the early evolution of the Universe and decrease gradually to constant values in the late time. The EoS parameter $ \omega $ varies in the range $ \left[\frac{-3A+n(2A+1)}{3A+n}, -1 \right] $. The role of the $ f(R,T) $ coupling constant bounds and the limit for the EoS parameter will be discussed in another  subsection.\\

We shall examine the behaviors of the physical and geometrical parameters in the following section more explicitly with the help of a graphical representation by expressing the cosmological parameters in terms of the redshift $ z $.
\begin{table}
\textbf{\caption{ Behavior of cosmological parameters}}
\begin{center}
\label{tabparm}
\begin{tabular}{l c c c c c c r} 
\hline\hline 
{Time ($ t $)}  \,\,\, & \,\,\,  $ a $   \,\,\, &\,\,\,  $ H $  \,\,\,&  \,\,\,  $ q $   \,\,\, & \,\,\,  $ \rho $   \,\,\,& \,\,\, \,\,\, $ p $  \,\,\, & \,\,\, \,\,\,  $ \omega $   \,\,\,\,\,\,&\,\,\,   $ \Lambda $ 
\\
\\
\hline 
\\
{$ t\rightarrow 0 $} & $ 0 $ & $ \infty $ & $ n-1 $ & $ \infty $ & $ \infty $ & $\frac{-3A+n(2A+1)}{3A+n}$ & $ \infty $ \\
\\
{$ t\rightarrow \infty $} & $ \infty $ & $ \alpha $ & $ -1 $ & $ \frac{3\alpha ^{2}}{(A+1)} $ & $\frac{-3\alpha ^{2}}{(A+1)} $ & $ -1 $ & $ \frac{3\alpha^{2}}{(A+1)} $  
\\
\\ 
\hline\hline  
\end{tabular}    
\end{center}
\end{table}
\section{Dynamics of the model}

\qquad In this study, we are trying to evaluate a mathematical cosmological model which can determine the dynamics of the Universe by explaining the behavior of its geometrical as well as physical parameters on large scale. There are around $4$ to $20$ cosmological parameters through which the dynamical behavior of the Universe can be quantified. Among these, the most fundamental cosmological parameters are the Hubble parameter $ H(t) $ and the deceleration parameter $ q(t) $. The other geometrical parameters can be determined by expanding the scale factor $ a(t) $ in the neighborhood of $ t_{0} $ by Taylor theorem as \cite{sin6}
\begin{equation}\label{10}
a(t)=a(t_{0}+t-t_{0})=a_{0}+\frac{(t-t_{0})}{1!}\dot{a_{0}}+\frac{(t-t_{0})^{2}}{2!}\ddot{a_{0}}+\frac{(t-t_{0})^{3}}{3!}\dddot{a_{0}}+\cdots ,
\end{equation}
where $ a_{0} $ represents the value of $ a(t) $ at the present time $ t_{0} $. The parameters $ H $ and $ q $ specify the significance of Einstein field equations and explain the recent astronomical observations accomplished by Eq. (\ref{10}). The involvement of higher derivative terms of the scale factor $ a(t) $ in Eq. (\ref{10}) extends the cosmographic analysis of the geometrical parameters \cite{vis1,vis2}. From Eq. (\ref{10}), one can define some geometrical parameters such as the jerk, snap, and lerk parameters, including the Hubble and deceleration parameters, through the higher derivatives of the scale factor as 
\begin{equation}\label{10a}
H=\frac{\dot{a}}{a},~~q=-\frac{\ddot{a}}{aH^{2}},~~j=\frac{\dddot{a}}{aH^{3}},~~s=\frac{\ddddot{a}}{aH^{4}},~~l=\frac{\ddddot{\dot{a}}}{aH^{5}}.
\end{equation}

In the following subsections, we discuss the behaviors of all these geometrical parameters for our model in detail. Moreover, we express the cosmological parameters in terms of the redshift ($ 1+z=\frac{a_{0}}{a} $) with normalized scale factor $ a_{0}=1 $. Here, we establish the $ t-z $ relationship, which turns out to be $ t(z)=\frac{1}{n\alpha }\log \left(1+(1+z)^{-n}\right) $. The Hubble parameter $ H $ which explains the dynamics of the Universe can be written in terms of the redshift as 
\begin{equation} \label{11a}
H(z)=\alpha (1+(1+z)^{n}), 
\end{equation}
or 
\begin{equation}\label{11b}
H(z)=\frac{H_{0}}{2}(1+(1+z)^{n}).  
\end{equation}
In the next subsection we will discuss the different phases of evolution of deceleration parameter with respect to redshift $z$ and examine the phase transition.

\subsection{Phase transition from deceleration to acceleration}

\qquad The deceleration parameter is examined as one of most influential cosmological parameters among the various cosmological parameters which describe the dynamics of the Universe. In this section, we discuss the different phases of the evolution of deceleration parameter. Cosmological observations indicate that the Universe experiences a cosmic speed up at late time implying that the Universe must have passed through a slower expansion phase in the past \cite{per,rie}. Moreover, a decelerating phase is also necessary for the formation of the structure. The cosmic transit from deceleration to acceleration or the \textit{phase transition} may be treated as a necessary phenomenon while describing the dynamics of the Universe. The above considered parametrization of the Hubble parameter in Eq. (\ref{9a}) which yields a time dependent expression of the deceleration parameter in Eq. (\ref{10d}) is rational with a phase transition. The present cosmic accelerating behavior can be estimated through the values of the deceleration parameter $q$ that belong to the negative domain. Keeping all these things in mind, we plot the graph of $ q $ with respect to the redshift $ z $ and choose the model parameter $ n $ suitably so that we have a phase transition redshift $ (z_{tr})$ exhibiting early deceleration to late acceleration. The deceleration parameter in terms of the redshift $ z $ can be written as

\begin{equation}\label{12}
q(z)=\frac{(n-1)(1+z)^{n}-1}{1+(1+z)^{n}}.  
\end{equation}

From this expression, we find the range of the deceleration parameter $q\in \lbrack (n-1),-1]$. As $ n>1 $, we see that lower limit is positive and the upper limit is negative, showing a signature flip. The present value of the deceleration parameter $ q $ is given by $ q_{0}=\frac{n}{2}-1 $ at $ z=0 $. Here, we are interested in examining the present era of the Universe as suggested by the observations \cite{per, rie, perc, ste}. Therefore, we assume the restriction $ 1<n<2 $ on our model parameter in such a way that a phase transition from early deceleration to present acceleration occurs. Thus, in this context, to have a negative value of the deceleration parameter, we have to choose the value of the model parameter n in such a range. Choosing the model parameter $ n $ suitably, $ q(z) $ can be plotted for a close view to discuss the behavior of the deceleration parameter as shown in Fig. 1.\\

\begin{figure}[tbph]
\begin{center}
$%
\begin{array}{c@{\hspace{.1in}}cc}
\includegraphics[width=3.5 in, height=2.5 in]{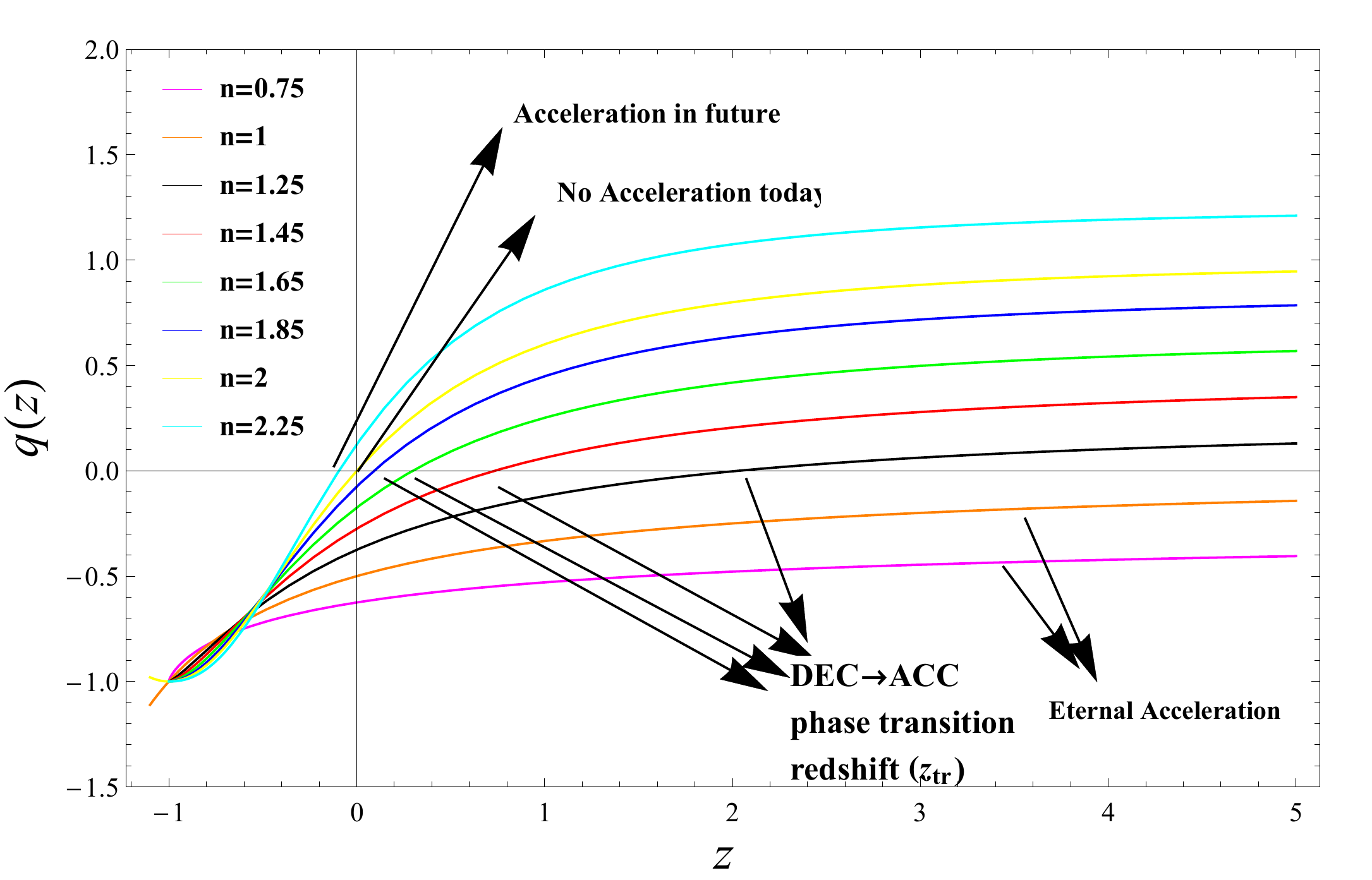} &  & 
\end{array}%
$%
\end{center}
\caption{\scriptsize The plots of deceleration parameter \textit{\ vs.} redshift $z$ for different $n$.}
\end{figure}

In Fig. 1, it is clearly observed that the deceleration parameter $ q $ is positive and negative for high and low redshift $ z $, respectively, in the range $ 1<n<2 $ of the model parameter $ n $. It has also been noticed explicitly that the phase transition from decelerating to accelerating regimes of the Universe depends on the variation of the model parameter $ n $. For the $n\leq 1$, model exhibits eternal acceleration, for $ n=2 $, the Universe shows no acceleration at present, and for $ n>2 $, acceleration of the Universe is possible in the near future. The plot shows the phase transition redshift $ (z_{tr}) $ for various values of $ n $ in the feasible range, $ n\in (1,2)$. For $n=1.25,$ $q=0$ at $z_{tr}=1.988$, for $n=1.45,$ $q=0$ at $z_{tr}=0.73$, for $n=1.65,$ $q=0$ at $z_{tr}=0.29$ and for $n=1.85,$ $q=0$ at $z_{tr}=0.091$. The present value of deceleration parameter $q_{0}$ corresponding to $n=1.25,$ $ n=1.45,$ $n=1.65$ and $n=1.85$ are $-0.371,-0.275,-0.179$ and $-0.077$ respectively. The best fit values of the model parameter $ n $ from various observations are discussed in Sect. 5.

\subsection{Physical significance of $ \lambda $ in the evolution of the Universe}

\qquad In order to explain the formation of structure, we know that, along with decelerated expansion, which is responsible for the structure formation one must require a kind of matter fluid that could produce a Jeans instability \cite{james}, and this is possible only when a low pressure fluid occurs in the Universe. To understand the structure formation in detail, a force of gravity is required which puts gas molecules together because as gravity pushes gas molecules closer, pressure and heat are produced, which then tends to push the molecules further apart. In $ 1902$ Jeans was the first person who calculated the region of influence mathematically, called the Jeans length \footnote {The Jeans length in a region can be  calculated by the formula $ L_J=\sqrt{\frac{\pi k T}{m G \rho}} $, where $ k $ is the Boltzmann's constant, $ T $ is the temperature of the gas, $ m $ is the mass of the atom in the gas, $ G $ is the gravitational constant and $\rho$ is the density of the gas. The Jeans length at the time of decoupling was approx. $108$ light years.} required to cause gravity to push atoms together and merge into structures like stars, galaxies or in fact global clusters.\\

In the present study, for structure formation we must have pressure $ p>0 $ in the early phase of the Universe and $ p<0 $ in the late phase, which could produce anti-gravitational effects to accelerate the Universe. This simply implies that the EoS parameter $ \omega $ must be positive in the early Universe and negative in the late Universe  with $ \omega=0 $ at a certain time of the cosmic evolution. For non-vanishing denominator of Eq. (\ref{omega}), we find a restriction on the coupling constant $\lambda$  as $ q\neq -(24\pi+4\lambda) $ for $ \lambda\neq 0 $. Moreover, $ \omega $ transits from early positive to late negative value, passing through $\omega=0$, which gives a relation for the deceleration parameter: 
\begin{equation}\label{omega0}
q=\frac{8\pi}{16\pi+3\lambda},
\end{equation} 
and using Eqs. (\ref{12}) and (\ref{omega0}), we have $\omega=0$ at redshift $z$ given by 
\begin{equation}\label{omega01}
z=\left[n\Big(\frac{16\pi+3\lambda}{24\pi+3\lambda}\Big)-1\right]^{\frac{-1}{n}}-1.
\end{equation}
	
From Eqs. (\ref{omega0}) and (\ref{omega01}), we see that the redshift $ z $, coupling constant $\lambda$ and the model parameter $ n $ are closely related. In this study we have chosen four particular values of $ n $, $ n=1.25, n=1.45, n=1.65 $ and $ n=1.85 $. We obtained the values of $\lambda$ as $-39.0954, -47.2191, -64.627$ and $-128.456$, corresponding to $n=1.25,1.45,1.65$ and $1.85$, respectively from the analysis of Eq. (\ref{omega01}) at present time $z=0$. By various observations, it is confirmed that our Universe is accelerating due to the presence of a mysterious form of energy known as dark energy, containing high negative pressure at the present, which produces a repulsive force. To achieve this kind of scenario in our study, it is essential to consider the value of the coupling constant $ \lambda $ in such a way that $ \omega $ changes its sign in the redshift range $ 0<z<1 $, which is consistent with observations. Therefore, we choose the values of $\lambda$ accordingly for all the four values of $ n $ and find a fixed particular value of $ \lambda $ as $ \lambda=-130 $. This value of $ \lambda $ is consistent with the current cosmic behaviors of the physical parameters $ \rho $, $ p $, $ \omega $ and $ \Lambda $. One may also consider a positive coupling constant $ \lambda $ accordingly such that $ \omega $ changes its sign from positive to negative but we find that, for some large positive values of $ \lambda $, $ \omega=0 $ does not seem to be consistent at high redshift ( $ z >1 $) and for small positive values of $ \lambda $, the model exhibits eternal acceleration, since $ \omega $ remains negative throughout the evolution ( $\omega<0 $). Therefore, we choose a suitable value of $ \lambda=-130 $ from a wide range of values of the coupling constant $ \lambda $ by examining numerically, which could meet the requirements of the current observations and we shall discuss a particular model with $ \lambda=-130 $ (see Fig. 2). The plot of the EoS parameter $ \omega $ vs. redshift $ z $ explains it well.\\ 

\begin{figure}[ht]
	\begin{center}
		$
		\begin{array}{c@{\hspace{.1in}}cc}
		\includegraphics[width=3.5in, height=3in]{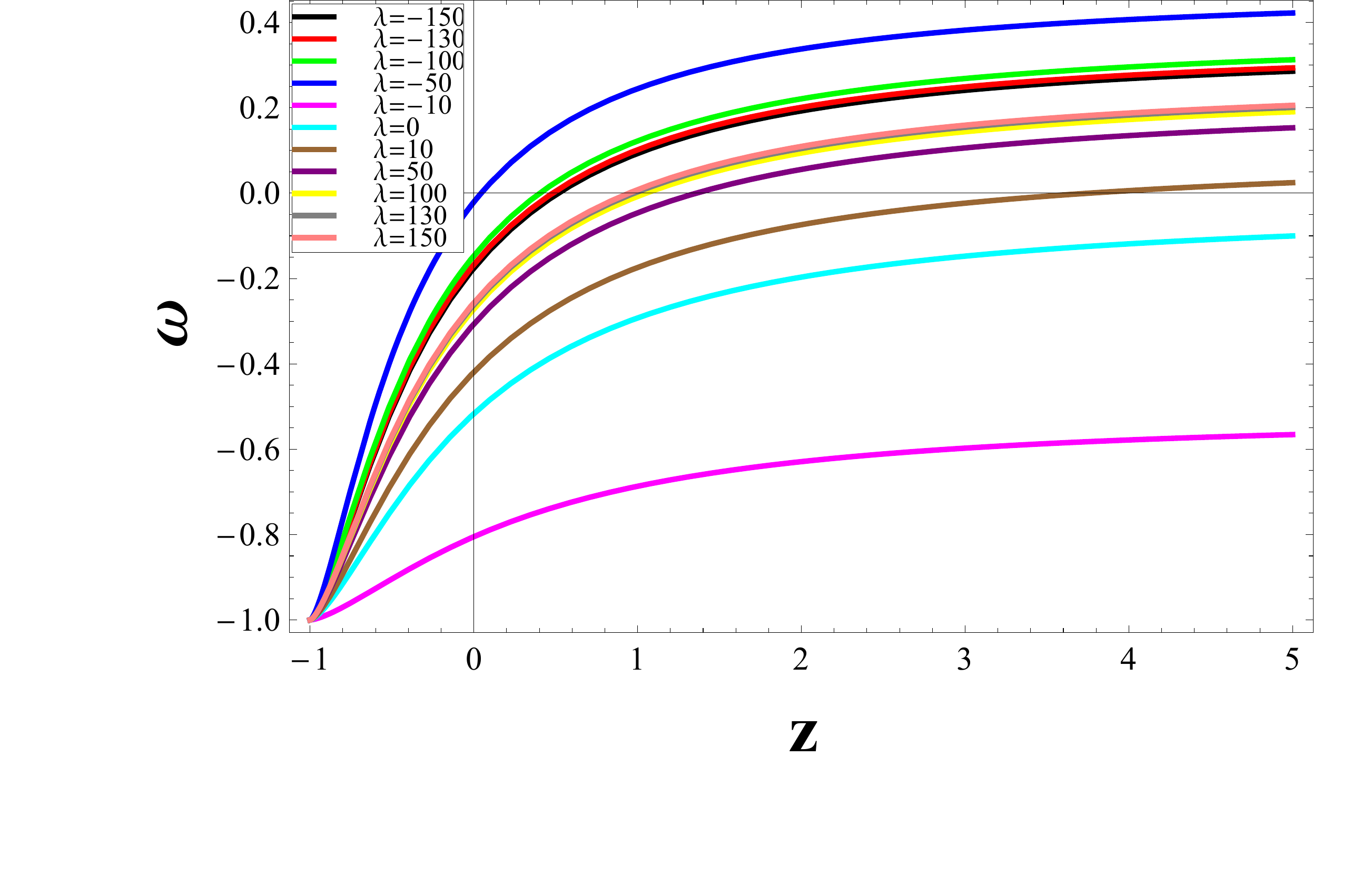} &  &  \\ 
		&  & 
		\end{array}
		$
	\end{center}
	\caption{\scriptsize The plot of $\protect\omega$ \textit{vs.} $z$ for $ n=1.45 $.} 
	\end{figure}
	
Here, we fix $ n=1.45 $ where the phase transition redshift is $ z_{tr}\approx 0.7 $. Now we observe the role of coupling constant $ \lambda $ played in the evolution of the EoS parameter $ \omega $ by taking different values of $ \lambda $. For $\lambda=0$, the case of GR and $\lambda=-10$, EoS parameter $\omega$ remains negative throughout the evolution. If we increase the value of $\lambda$ from a small positive number to a large positive number, the redshift transition time gets shifted from right to left. Also if we take high negative values of $\lambda$ from $-50$ to $-150$, the redshift phase transition time gets shifted from left to right and ultimately remains in the interval $z\in (0,1)$. For all the cases of $\lambda$ plotted here other than $\lambda=0, -10$, it is important to notice that the matter content in the Universe behaves like perfect fluid in the initial phase of the Universe and later on the Universe enters into a quintessence regime and finally approaches $\omega=-1$ as $z\to -1$ but it never crosses the phantom divide line. Hence, the coupling constant $ \lambda $ has great importance in the reconstruction of the cosmic evolution of our model. 

\subsection{Physical parameters and their evolution}

\qquad In the following section, we analyze the evolution of the energy density $ \rho $, isotropic pressure $ p $, EoS parameter $ \omega $ and the cosmological constant $ \Lambda $ for our model from the parametrization (\ref{9a}). Using $ t-z $ relationship, we get the expressions for $ \rho $, $ p $, $ \omega $ and $ \Lambda $ in terms of redshift as follows:

\begin{equation}\label{rhoz}
\frac{\rho (z)}{H_{0}^{2}}=\frac{3\lambda}{4(8\pi+2\lambda)}\left[(1+(1+z)^{n})^2\right] \left[\frac{\lambda^2}{(8\pi+\lambda)(8\pi+2\lambda)} n (1+z)^n (1+(1+z)^n)\right],  
\end{equation}

\begin{equation}\label{pz}
\frac{p(z)}{H_{0}^{2}}=-\frac{3\lambda}{4(8\pi+2\lambda)}\left[(1+(1+z)^{n})^2\right]+\frac{\lambda (16\pi+3\lambda)}{(8\pi+\lambda)(8\pi+2\lambda)} n (1+z)^n \left[(1+(1+z)^n)\right],  
\end{equation}

\begin{equation}\label{wz}
\omega (z)=\frac{-3(8\pi+\lambda)
(1+(1+z)^{-n})+n(16\pi+3\lambda)}{3(8\pi+\lambda)(1+(1+z)^{-n})+n \lambda},  
\end{equation}

\begin{equation}\label{Lz}
\frac{\Lambda (z)}{H_{0}^{2}}=\frac{3\lambda}{(8\pi+2\lambda)}\left[1+(1+z)^n\right]^2-(\frac{\lambda}{(8\pi+2\lambda)})n(1+z)^n \left[1+(1+z)^n\right].  
\end{equation}

The evolution of the physical parameters in Eqs. (\ref{rhoz})-(\ref{Lz}) are shown in the figures.

\begin{figure}[tbph]
\begin{center}
$%
\begin{array}{c@{\hspace{.1in}}cc}
\includegraphics[width=2.9 in, height=2.5 in]{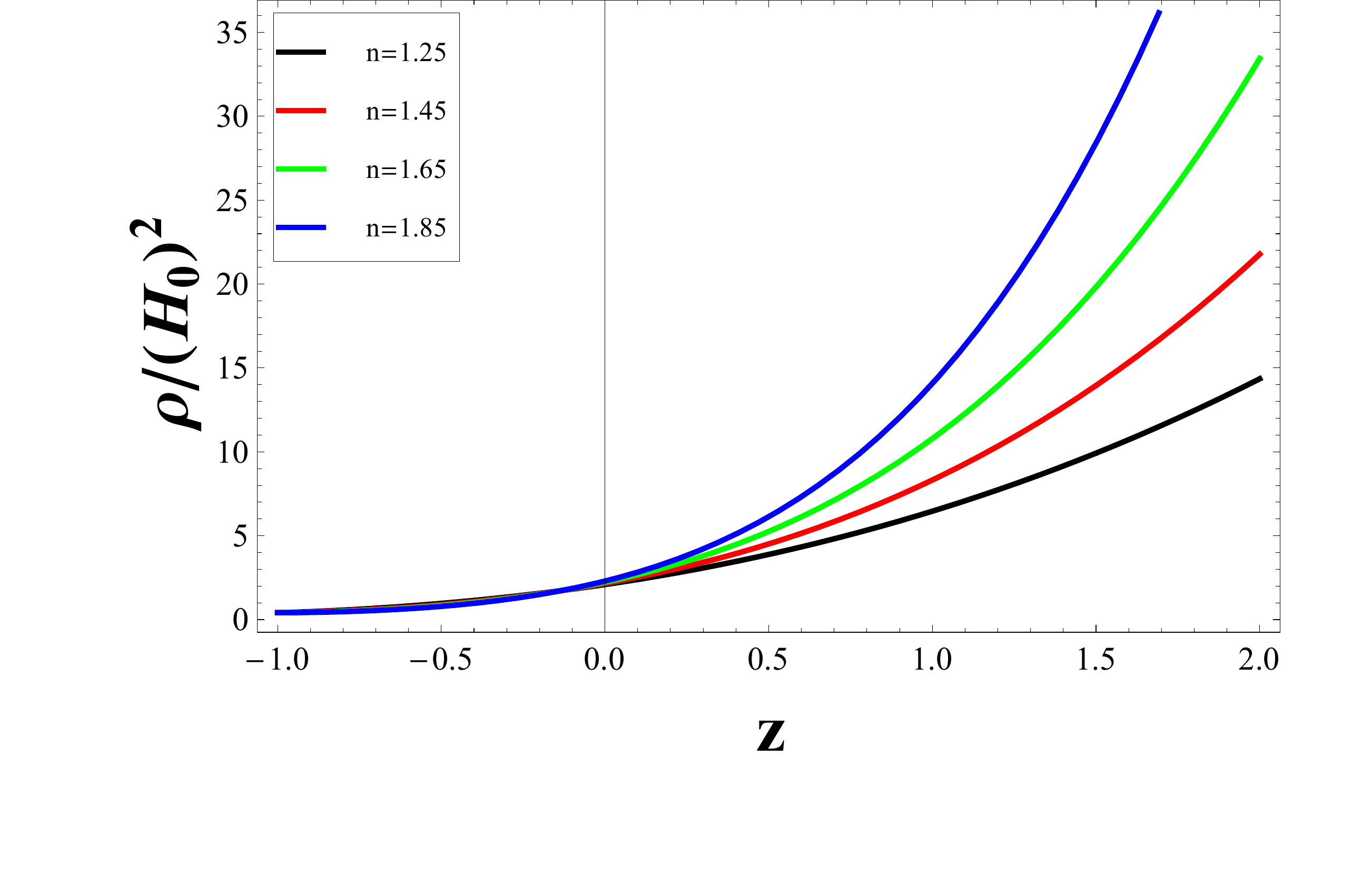} & 
\includegraphics[width=2.9 in, height=2.5 in]{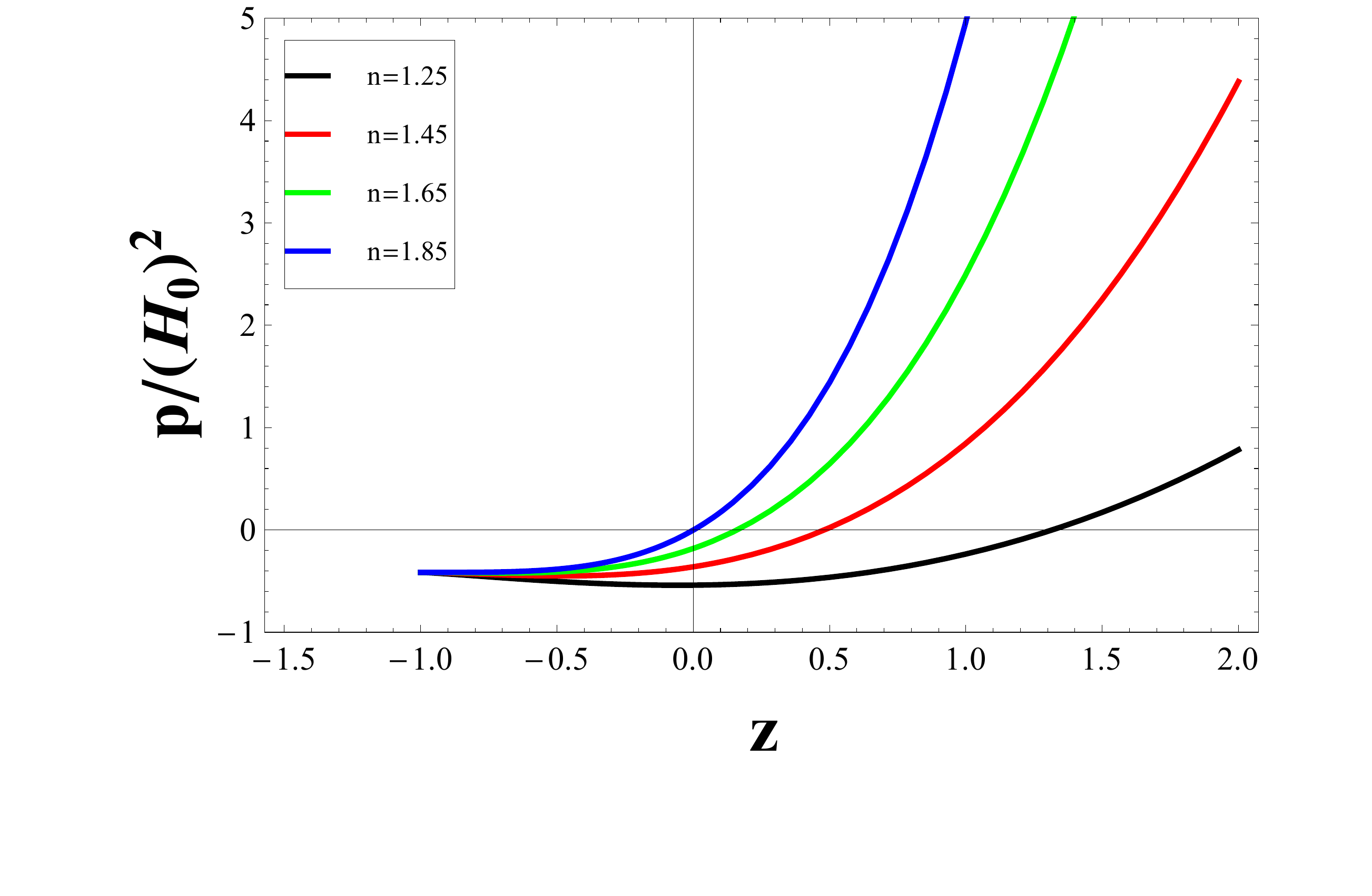} & \\ 
\mbox (a) & \mbox (b) & 
\end{array}%
$%
\end{center}
\caption{\scriptsize The plots of the energy density $\protect\rho $ and pressure $ p $ \textit{ vs.} redshift $z$ with $\protect\lambda =-130$.}
\end{figure}

Figure 3a depicts the evolution of the energy density $\rho$ with respect to the redshift $ z $ for different values of model parameter $ n $ mentioned in the plot. For a high redshift, the energy density is very high, as expected; then energy density falls as time unfolds and later on it approaches $\frac{3\alpha ^{2}}{(A+1)}$ as $ z\rightarrow -1 $. \\
 
Figure 3b highlights the picture of the isotropic pressure for the specified values of $ n $. In the initial phases of the early Universe for a very high redshift, the isotropic pressure $p$ attains a very large value and approaches $\frac{-3\alpha ^{2}}{(A+1)}$ in the future as $ z\rightarrow -1 $. The negative values of cosmic pressure are corresponding to the cosmic acceleration according to the standard cosmology. Hence our model shows accelerated expansion at present as well as in the future evolution. We know that, discussing the primordial nucleosynthesis in the early Universe in a model, it is obvious that a deviation of more than a $ 10\% $ in the expansion rate with respect to the standard model during the nucleosynthesis era conflicts with the observed $ He^4_2 $ abundance. Therefore, according to our model, in the early Universe at high redshift, the matter all throughout the Universe was fairly dense so that regions of about $100$ light years (the Jeans length) across matter would coalesce and form global clusters corresponding to a high positive pressure. Therefore, our present model is in good agreement with the fact of the structure formation in the Universe.\\

\begin{figure}[tbph]
	\begin{center}
		$%
		\begin{array}{c@{\hspace{.1in}}cc}
		\includegraphics[width=2.9 in, height=2.5 in]{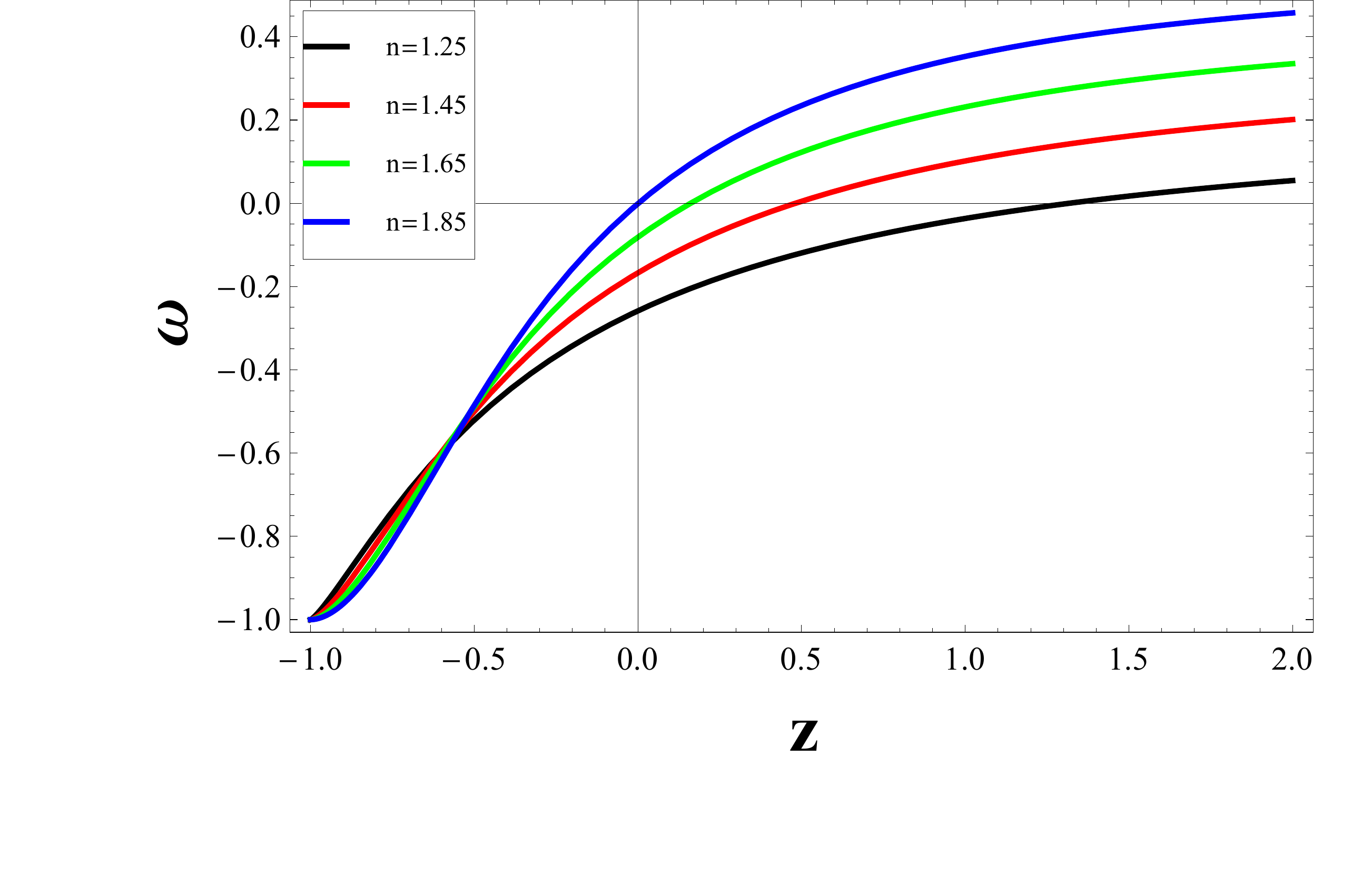} & 
		\includegraphics[width=2.9 in, height=2.5 in]{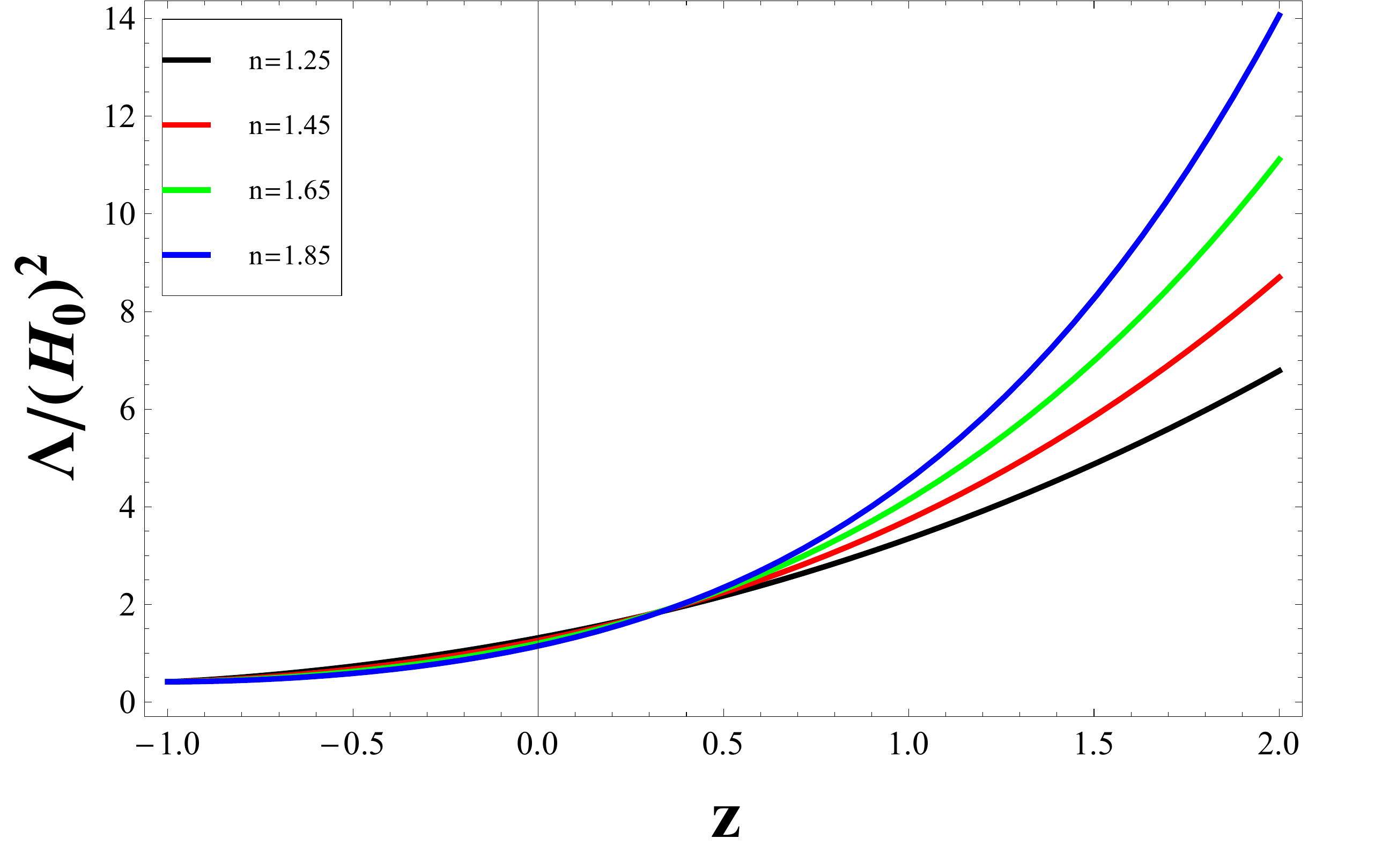} & \\ 
		\mbox (a) & \mbox (b) & 
		\end{array}%
		$%
	\end{center}
	\caption{\scriptsize The plots of the EoS parameter $\protect\omega $ and cosmological constant $\Lambda $ \textit{\ vs.} redshift $z$ with $\protect \lambda =-130$.}
\end{figure}

The profile of the EoS parameter $\omega $ and cosmological constant $\frac{\Lambda }{H_{0}^{2}} $ is investigated in Fig. 4. The behavior of $\omega $ with respect to the redshift $z$ can be seen in Fig. 4a. For all values of the model parameter $n$ and a fixed value of the coupling constant $\lambda=-130 $, $\omega$ takes positive values in the early Universe; then $\omega$ starts changing its sign from positive to negative and $\omega=0$ at $z=1.31678, 0.481009, 1.060876$ and $z=0.001195$ corresponding to $n=1.25,1.45,1.65$ and $n=1.85$, respectively. After this, $\omega$ enters the quintessence region, and $\omega \rightarrow -1$, in late time as $z\rightarrow -1$ which suggests that matter in the Universe behaves like a perfect fluid initially. Later on the model is similar to a dark energy model and behaves like a quintessence model and finally approaches −1 	without entering the phantom region. Figure 4b depicts the variation of the cosmological constant $\left( \frac{\Lambda }{H_{0}^{2}}\right) $ with respect to the redshift $z$. It has been observed that the cosmological constant remains positive throughout the cosmic evolution, decreasing in nature and reaching a small positive value at present epoch $ z\rightarrow 0 $, favoring the observations \cite{per,rie,ton,clo} and $\Lambda \rightarrow \frac{3\alpha ^{2}}{(A+1)}$ as $z\rightarrow -1$. The outcome from these observations suggests a very minute positive value having magnitude $ \sim10^{-123} $.

\subsection{Physical significance of jerk, snap, lerk parameters}

\qquad For our model, the expressions for jerk parameter $ j $, snap parameter $ s $ and lerk parameter $ l $ are obtained in terms of redshift $ z $, given by 
\begin{equation}
j(z)=(1+n(n-3)[(1+(1+z)^{-n})^{-1}+n^{2}(1+(1+z)^{-n})^{-2}],  \label{j}
\end{equation}%
\begin{eqnarray}
s(z) &=&1+n[-n^{2}(1+(1+z)^{-n})^{-3}-n(4n-7)(1+(1+z)^{-n})^{-2}  \notag \\
&&-(6+n(n-4))(1+(1+z)^{-n})^{-1}],  \label{snap}
\end{eqnarray}%
\begin{eqnarray}  \label{lerk}
l(z) &=&[1+n^{4}(1+(1+z)^{-n})^{-4}+n^{3}(11n-15)(1+(1+z)^{-n})^{-3}  \notag
\\
&&+n^{2}(25+n(11n-30))(1+(1+z)^{-n})^{-2}n(-10+n(10+n(n-5)))(1+(1+z)^{-n})^{-1}].
\end{eqnarray}

\begin{figure}[tbph]
\begin{center}
$%
\begin{array}{c@{\hspace{.1in}}cc}
\includegraphics[width=2.2 in, height=2.2 in]{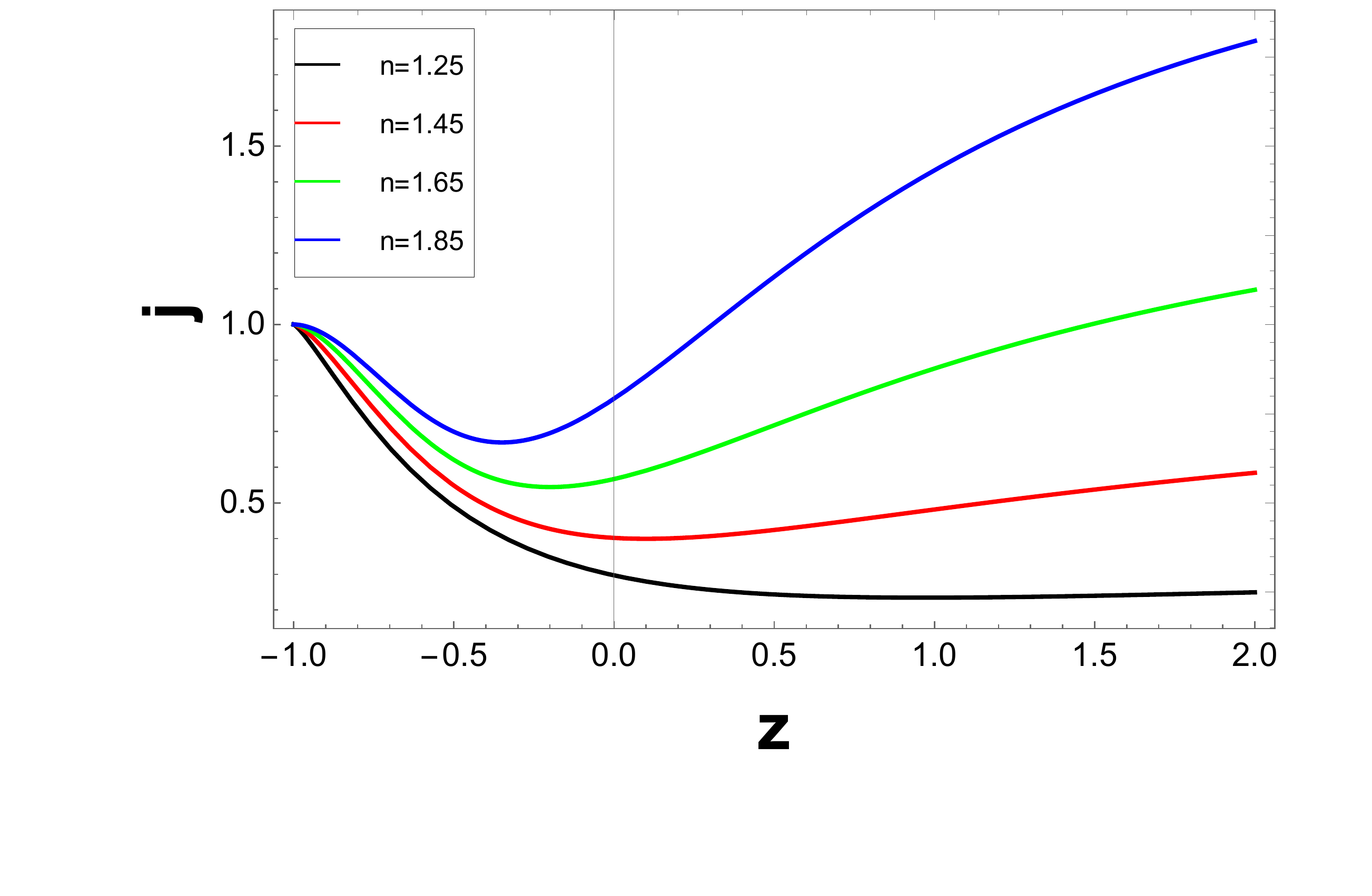} & %
\includegraphics[width=2.2 in, height=2.2 in]{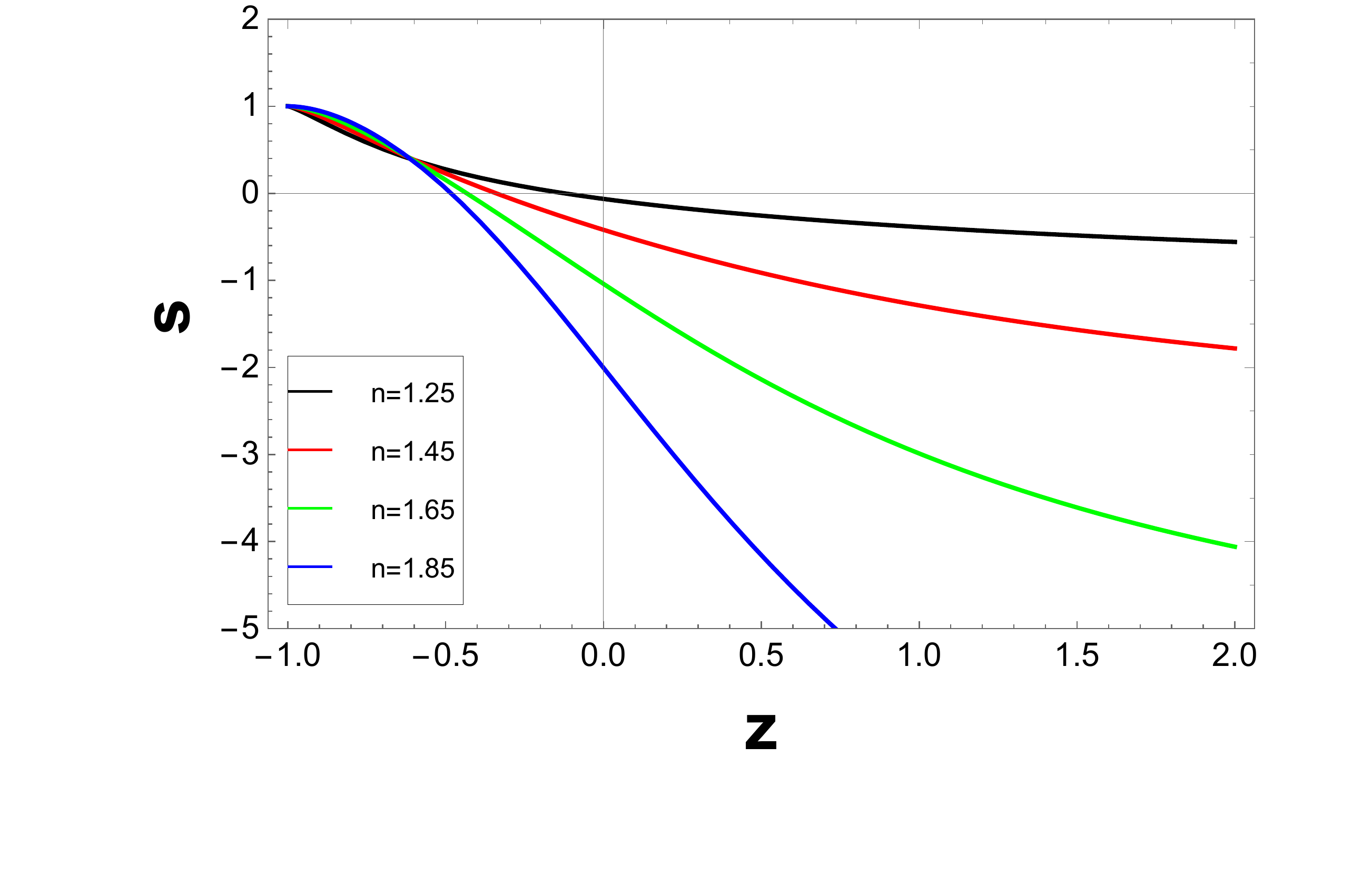} & %
\includegraphics[width=2.2 in, height=2.2 in]{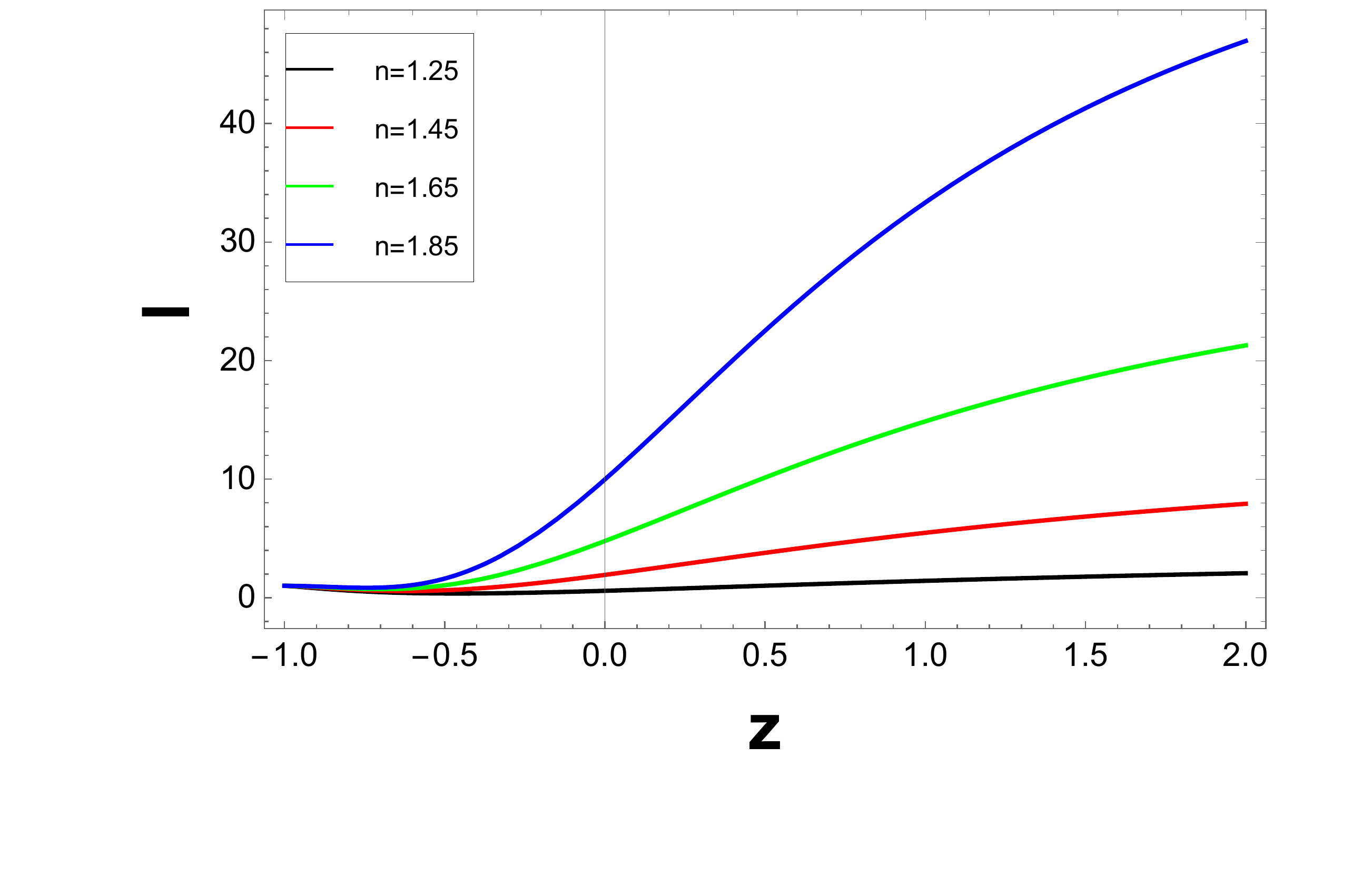} \\ 
\mbox (a) & \mbox (b) & \mbox (c)%
\end{array}%
$%
\end{center}
\caption{\scriptsize The plots of jerk $j$, snap $s$ and lerk $l$ parameters \textit{\ vs.} redshift $z$.}
\end{figure}
Figure 5 depicts the evolution of jerk $j$, snap $s$ and lerk parameter $l$  with respect to the redshift $z$. In Fig. 5(a), the evolution of the jerk parameter is represented for all the four values of $n$, and it can be observed that the $j$ parameter lies in the positive range throughout its course. Also $j \to 1$ as $z \to -1 $, $
\forall n$, which is consistent with observations of standard $ \Lambda CDM $ but at present $z=0$, $j\neq1$, $\forall n$. Therefore, our model is similar to the dark energy model, different from, $ \Lambda CDM $, $\forall n$, at the present time. Figure 5(b) enacts the profile of snap parameter $s$ during its evolution. In the early Universe $s$ assume value in the negative range $ \forall n$ then as Universe evolves, $s$ take values in the positive range, \textit{i.e.} in the entire evolution of $s$, there is one transition from negative to positive range. Also it can be directly seen from Fig. 5b that the transition of $s$ depends on the model parameter $n$, \textit{i.e.}, the transition redshift of $s$ is delayed as $n$ takes values from $1.25$ to $ 1.85 $. Fig. 5c shows the variation of the lerk parameter $l$ over the redshift $ z $. The lerk parameter $l$ assumes only positive values without any redshift transition. In addition to $j$, both $s$ and $l$ also approaches $1$ as in late time $ z\to -1$.

\section{Physical analysis and geometrical diagnostic}

\subsection{Energy conditions}

\qquad In the general theory of relativity, energy conditions (ECs) have great advantage for broad understanding of the singularity theorem of space-time. ECs are considered as the basic ingredient to describe the role of different geodesics \textit{i.e.,} null geodesics, space-like, time-like or light-like geodesics. The additional privilege of EC is to provide the elementary tool for study certain ideas as regards black holes and worm holes. There are several ways in which ECs can be formulated, \textit{e.g.} geometric way,
physical way or in effective way. The viability of various types of point wise EC could be discussed by the well known Raychaudhuri equation \cite{carr}. The situation of exploring ECs in GR is to relate cosmological geometry with general energy momentum tensor in such a way that the energy remains positive \cite{mvis}. But generally this is not the case in modified gravity theories. Therefore, one has to be concerned while expressing such a relation in modified gravity. For the literature review of ECs have already been examined in the general theory of relativity, see \cite{san1,san2,sen}. Several issues in exploring the ideas of ECs have been proposed in modified gravity also. For a brief and recent reviews see \cite{san3,bert} in $f(R)$ gravity and  \cite{nojs,gar,bani} in $ f(G) $ gravity. The expressions for four types of EC in $f(R,T)$ gravity with effective energy density $\rho$ and isotropic pressure $p$ can be represented as follows:

\begin{itemize}
\item NEC $\Leftrightarrow$ $\rho+p_i \geq 0$, $ \forall i $,
\item WEC $\Leftrightarrow$ $\rho \geq 0$, $ \rho+p_i \geq 0 $, $ \forall i $,
\item SEC $\Leftrightarrow$ $\rho+\sum_{i=1}^3 p_i \geq 0$, $\rho+p_i \geq 0$, $ \forall i $,
\item DEC $\Leftrightarrow$ $ \rho \geq 0$ , $|p_i| \leq \rho $, $ \forall i $, $ i=1,2,3  $.
\end{itemize}

Also, if the energy density $ \rho $ and isotropic pressure $ p $ are described in terms of scalar field $ \phi $ (real), then energy conditions in terms of scalar field $ \phi $ satisfies:

\begin{itemize}
\item NEC: $\forall \, V(\phi)$,
\item WEC $\Leftrightarrow$ $V(\phi) \geq \frac{\dot{\phi}^2}{2}$,
\item SEC $\Leftrightarrow$ $V(\phi) \leq \dot{\phi}^2$,
\item DEC $\Leftrightarrow$ $V(\phi) \geq 0$.
\end{itemize}

\begin{figure}[tbph]
\begin{center}
$%
\begin{array}{c@{\hspace{.1in}}cc}
\includegraphics[width=2.2 in, height=2.2 in]{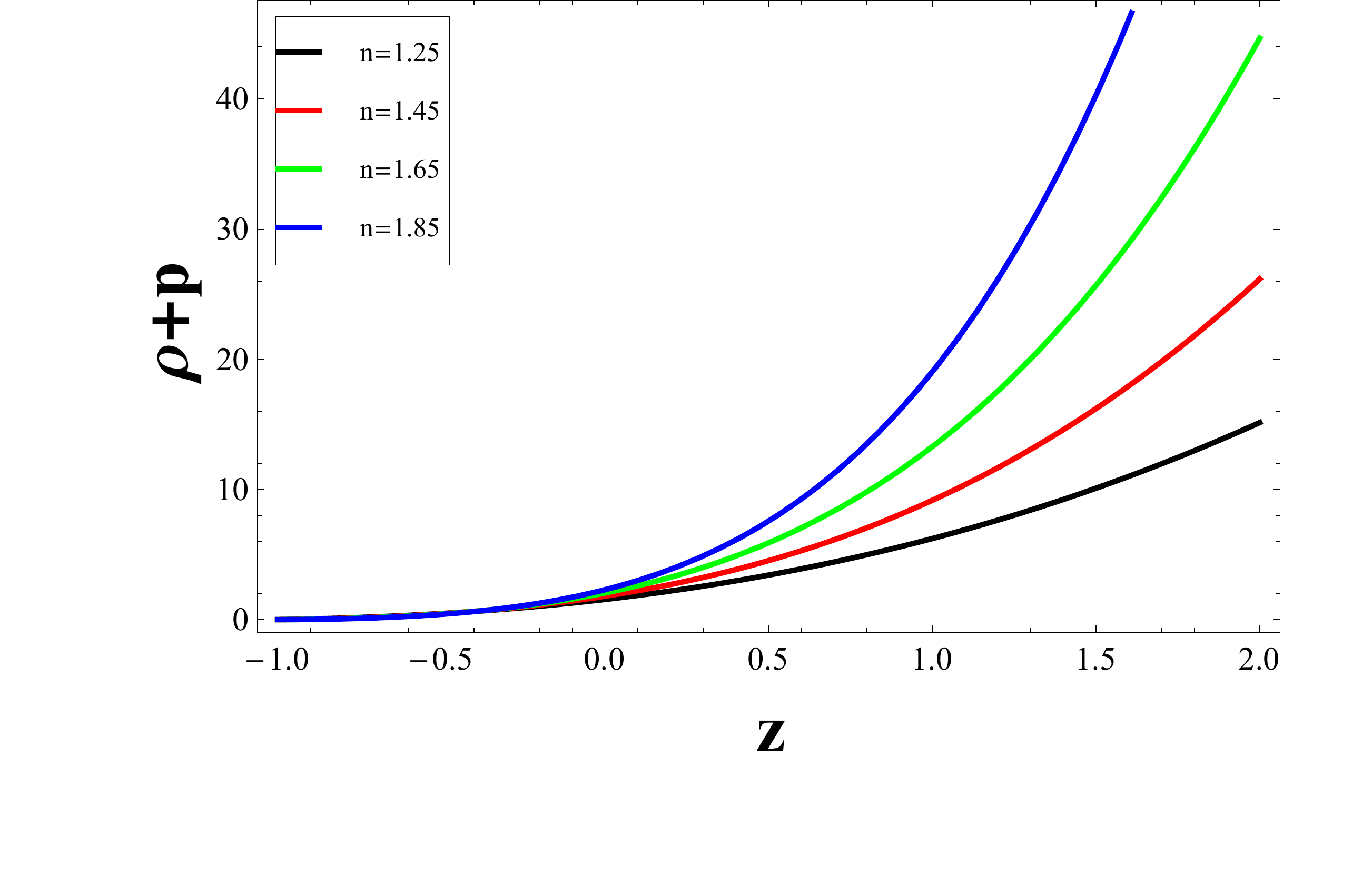} & %
\includegraphics[width=2.2 in, height=2.2 in]{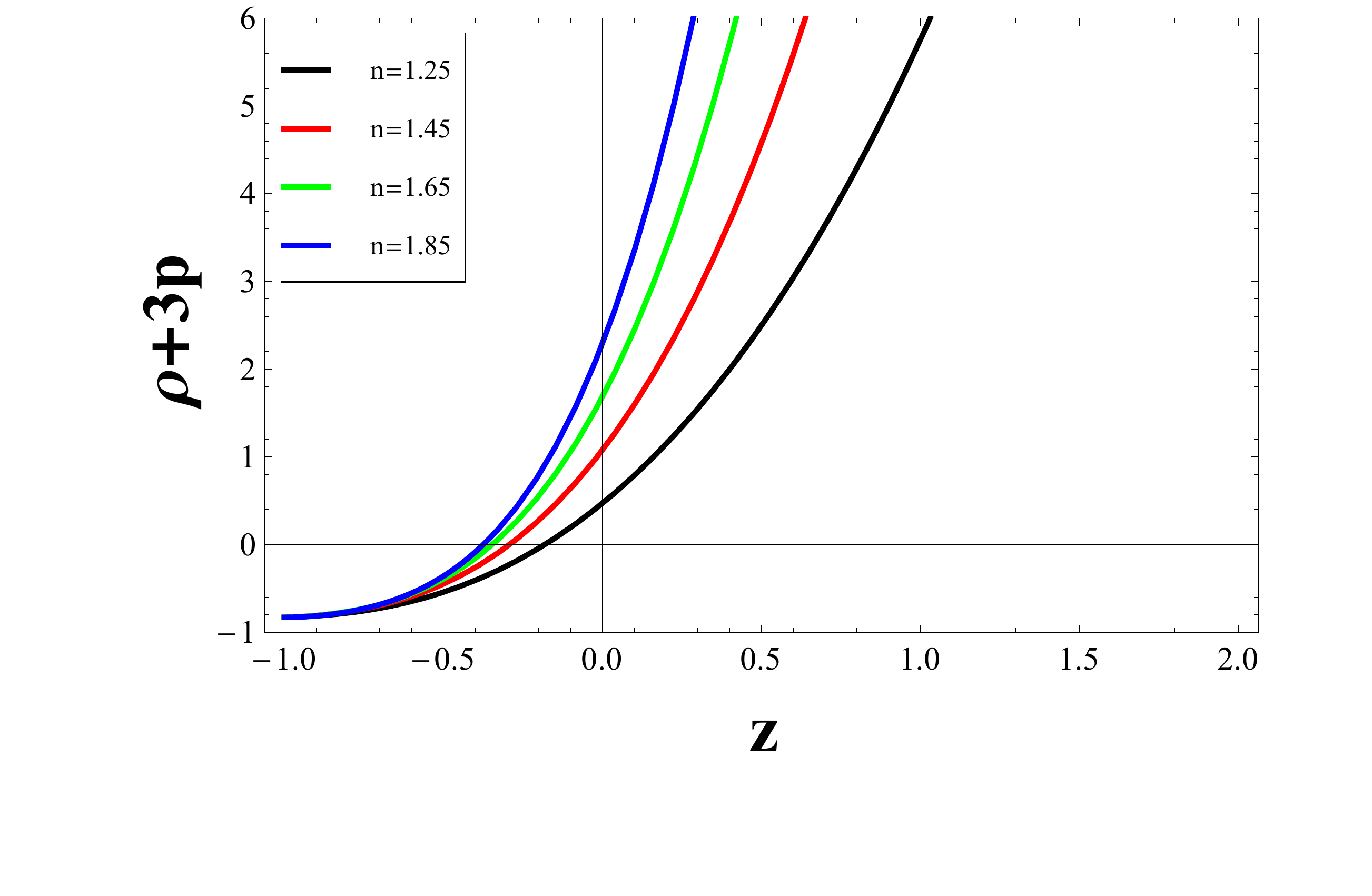} & %
\includegraphics[width=2.2 in, height=2.2 in]{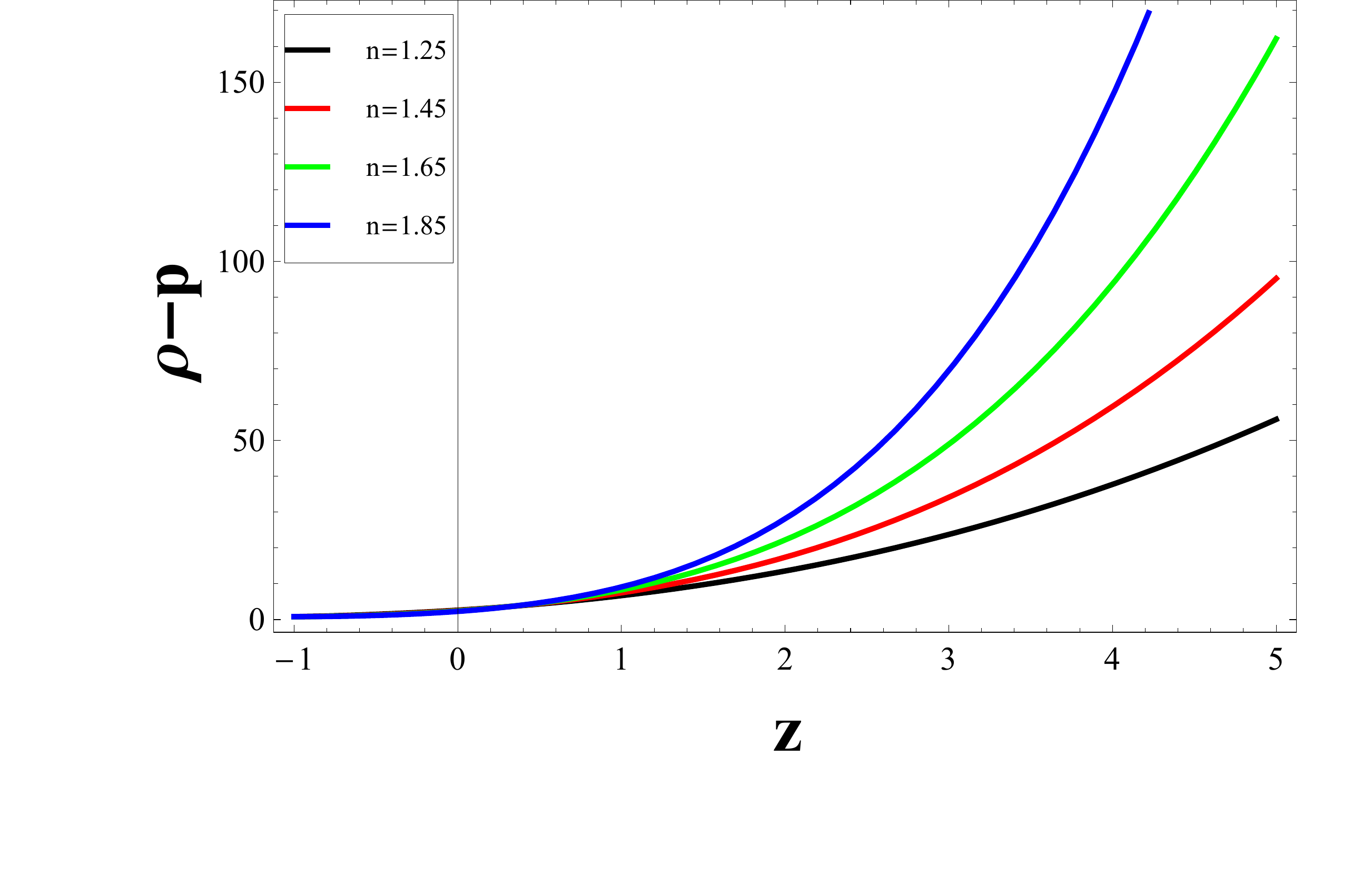} \\ 
\mbox (a) & \mbox (b) & \mbox (c)%
\end{array}%
$%
\end{center}
\caption{\scriptsize The plots of NEC, SEC and DEC for the model $ \lambda=-130 $.}
\end{figure}
Here, NEC, WEC, SEC and DEC are defined as the null energy condition, weak energy condition, strong energy condition and dominant energy condition, respectively.\\
Here, we present the graphs of NEC, SEC and DEC for all four values of the model parameter $ n $ and a fixed value of $f(R,T)$ coupling constant $\lambda$. We observe that, from Fig.6a-c, NEC and DEC hold for all values of $n$, while SEC fails for all values of $n$. 

\subsection{Statefinder diagnostic}
\qquad As is well known, the role of geometric parameters has great importance in order to study the dynamics of a cosmological model. In what follows, in Sect. $4.2$, we have discussed the different phases of the evolution of the deceleration parameter and concluded that the deceleration parameter alters its sign from positive to negative, corresponding to high redshift to low redshift, respectively. The phase transition of deceleration parameter provides hope to discover the source of recent the acceleration. Through the requirement of a more general dark energy model other than $ \Lambda CDM $ and the development in the accuracy of current cosmological observational data, there arises the problem of looking into the quantities involving higher derivatives of the scale factor $a$. \\

In order to have a general study of different dark energy models, a geometrical parameter pair technique, known as statefinder diagnostic (SFD), has been proposed \cite{var1,ala}; the pair are denoted by $\{r,s\} $, where $ r $ and $ s $ are defined as

\begin{equation}
r=\frac{\dddot{a}}{aH^{3}}\text{, \ \ }s=\frac{r-1}{3(q-\frac{1}{2})},
\end{equation}
where $q\neq \frac{1}{2}$.\\

Various dark energy scenarios can be examined by the distinct evolutionary trajectories of the geometric pair $\{r,s\}$ emerging in the $r-s$ plane in Fig. 7a. A symbolic feature of the SFD is that the standard $ \Lambda CDM $ model of cosmology is represented by the pegged point $\{r,s\}=\{1,0\}$, whereas the standard matter dominated Universe, $SCDM$, corresponds to the fixed point $\{r,s\}=\{1,1\}$. Other than the $ \Lambda CDM $ and $SCDM$ model, the SFD analysis can successfully discriminate among the several dark energy candidates such as quintessence, braneworld dark energy models, Chaplygin gas and some other interacting dark energy models by locating some particular region in the said diagram in the distinctive trajectories \cite{sin2, sami1, msha, sarita}.\\

Now we implement the SFD approach in our dark energy model to discuss the behavior of our model and study its converging and diverging nature with respect to the $SCDM$ and $ \Lambda CDM $ model. The expression for the $ r,s $ parameters for our model are

\begin{equation}
r=1+n[e^{-2 n \alpha t}(n-3)+n],
\end{equation}
and 
\begin{equation}
s=\frac{2n[3-n(1+e^{-n \alpha t})]}{9 e^{n \alpha t}-6n}.
\end{equation}

\begin{figure}[tbph]
\begin{center}
$%
\begin{array}{c@{\hspace{0.1in}}cc}
\includegraphics[width=2.4 in]{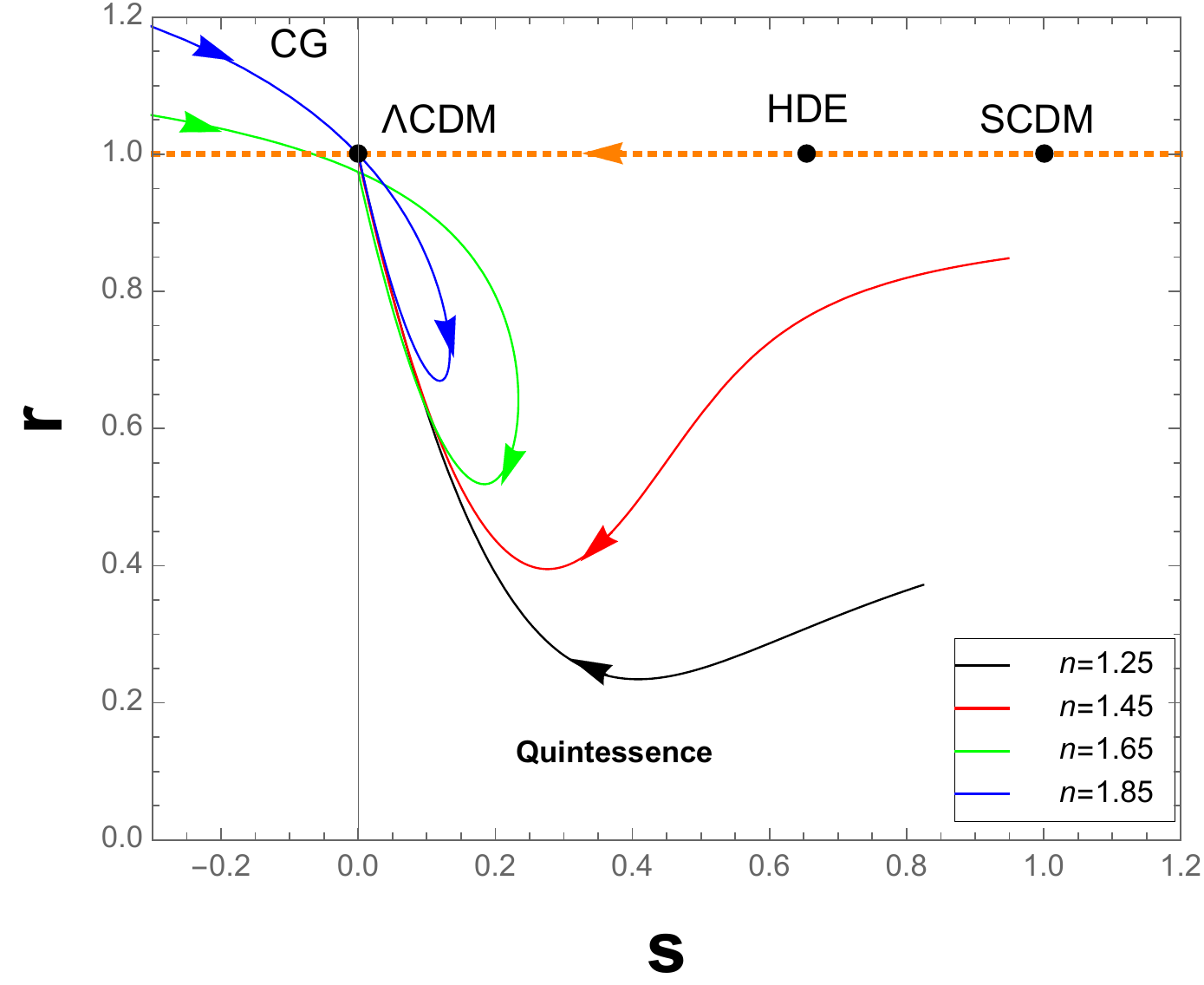} & 
\includegraphics[width=2.5 in,
height=2.01 in]{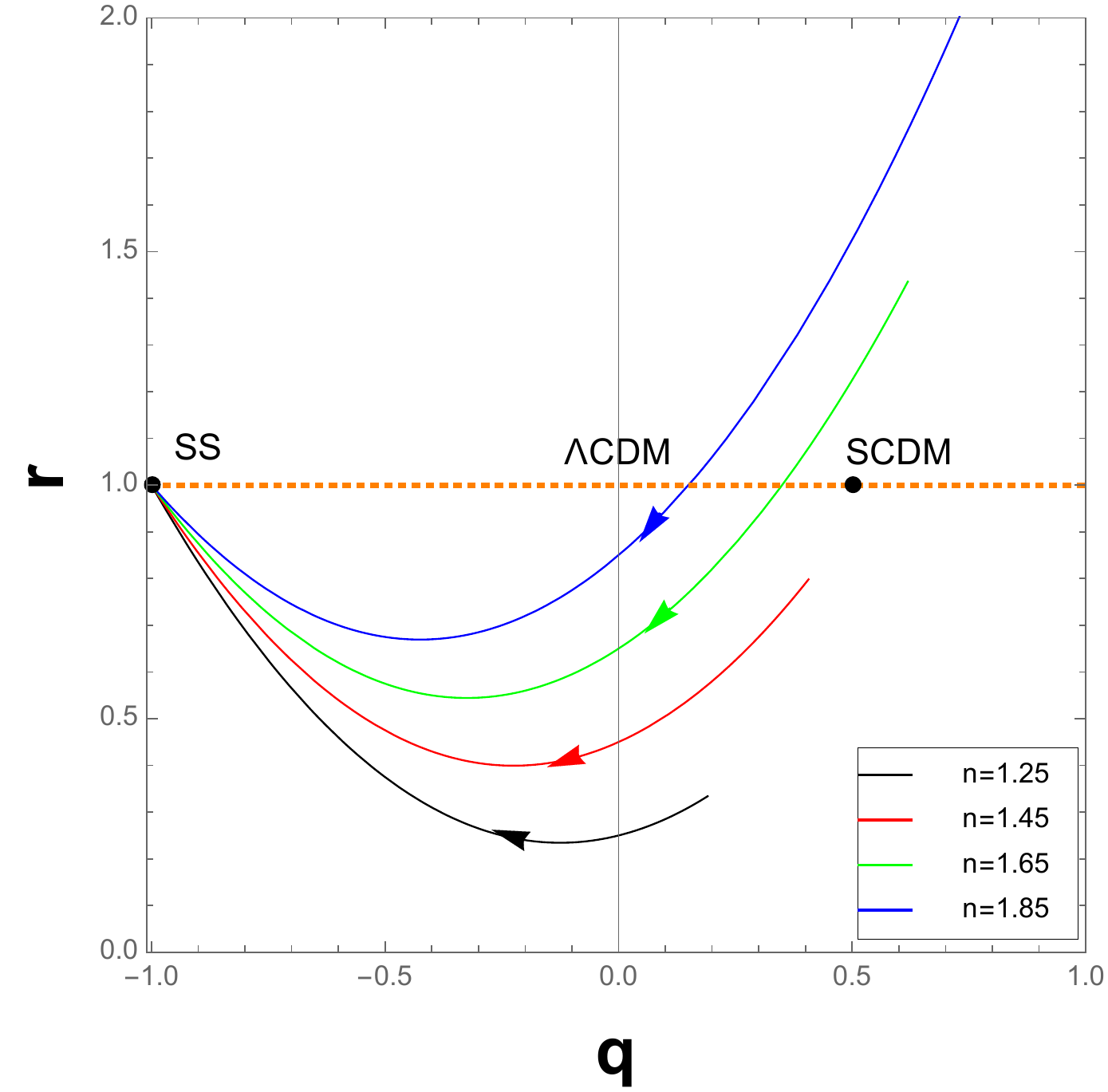} &  \\ 
\mbox (a) & \mbox (b) & 
\end{array}%
$%
\end{center}
\caption{\scriptsize The $ \mathit{s-r} $ and $ \mathit{q-r} $ diagrams for our model.}
\end{figure}

Figure 7a represents the time evolution of four trajectories for different values of n in $ r-s $ plane. All the trajectories corresponding to different values of $ n $ evolve with time but deviate from the point $ SCDM $, \textit{i.e.} $ \{r,s\}=\{1,1\} $ which corresponds to a matter dominated Universe. The directions of $ r-s $ trajectories in the plane diagram are represented by the arrows. Initially, we have examined that corresponding to $ n=1.25 $ and $ n=1.45 $; the trajectories remain in domain $ r<1, s>0 $, which relate our dark energy model to the quintessence model. Also trajectories corresponding to $n=1.65$ and $n=1.85$ start evolving from the region $ r>1,s<0$, which resembles the behavior of dark energy with Chaplygin gas and this region is highlighted by $CG$ in the top leftmost part of the plot. The downward pattern of trajectories  representing the CG behavior and the upward trend of the trajectories representing quintessence behavior are eventually met at the point $\{r,s\}=\{1,0\}$, \textit{i.e.}, we have the $ \Lambda CDM $ model. This suggests that our model behaves like $ \Lambda CDM $ in the late time of cosmic evolution. In addition we have presented one more horizontal line in the above diagram, which shows the transformation of trajectories from a matter dominated Universe $SCDM$ to $ \Lambda CDM $ as time unfolds. The point having coordinates $\{r,s\}=\{1,\frac{2}{3}\}$ on the horizontal line represents the holographic dark energy model with future event horizon as IR cut-off labeled as $HDE$ in Fig. 7a, begins the evolution from the point $ \{r,s\}=\{1,\frac{2}{3}\}$ and ultimately ends its evolution at $ \Lambda CDM $ \cite{mrs,mghu,djli}. Therefore, the plot of $\{r,s\}$ for our model is effectively discriminant among other dark energy model for different $ n $.\\

Figure 7b represents the time evolution of four trajectories for different values of n in $ r-q $ plane. Since we have seen the complete description of the phase transition of deceleration parameter in Sect. 3.2, we can again observe the phase transition of our model by looking into the trajectories of $r-q$ diagram as $ q $ changes its sign from positive to negative. The evolution of the trajectories for different values of $ n $, commences in the vicinity of a matter dominated Universe $ SCDM $ but never converges to $ SCDM $. As time evolves, the values of $r$ and $q$ start decline and they attain their minimum position, after which both $r$ and $q$ start to increase towards $ SS $, which is located in the diagram at $ (1,-1) $. The progression of the trajectories to $ SS $ suggests that our dark energy model may behave like the steady state model in late-time.

\subsection{Om diagnostic}

\qquad In this section, we use one more technique to differentiate the standard $ \Lambda CDM $ model from other dark energy models. This approach has been developed to examine the dynamics of the dark energy models by connecting the geometric parameter $H$ with redshift $z$, and it is known as Om diagnostic \cite{sha,zun, mshah1}. It is worth mentioning that Om diagnostic can make distinction among various dark energy models without actually referring to the exact present value of density parameter of matter and without comprising  the EoS parameter. Also Om diagnostic yields a null test for the cosmological constant $\Lambda $ as Om takes same constant value irrespective of the redshift z for $ \Lambda CDM $, which exhibits the non evolving behavior of Om when dark energy is a cosmological constant. Also Om diagnostic is a single parameter evaluation technique; therefore it is quite simple to formulate, as compared to SFD. Om diagnostic is defined as 
\begin{equation}
Om(z)=\frac{\left( \frac{H(z)}{H_{0}}\right) ^{2}-1}{z(z^{2}+3z+3)}.
\end{equation}%

\begin{figure}[tbph]
\begin{center}
$
\begin{array}{c@{\hspace{0.1in}}cc}
\includegraphics[width=2.9 in, height=2.4 in]{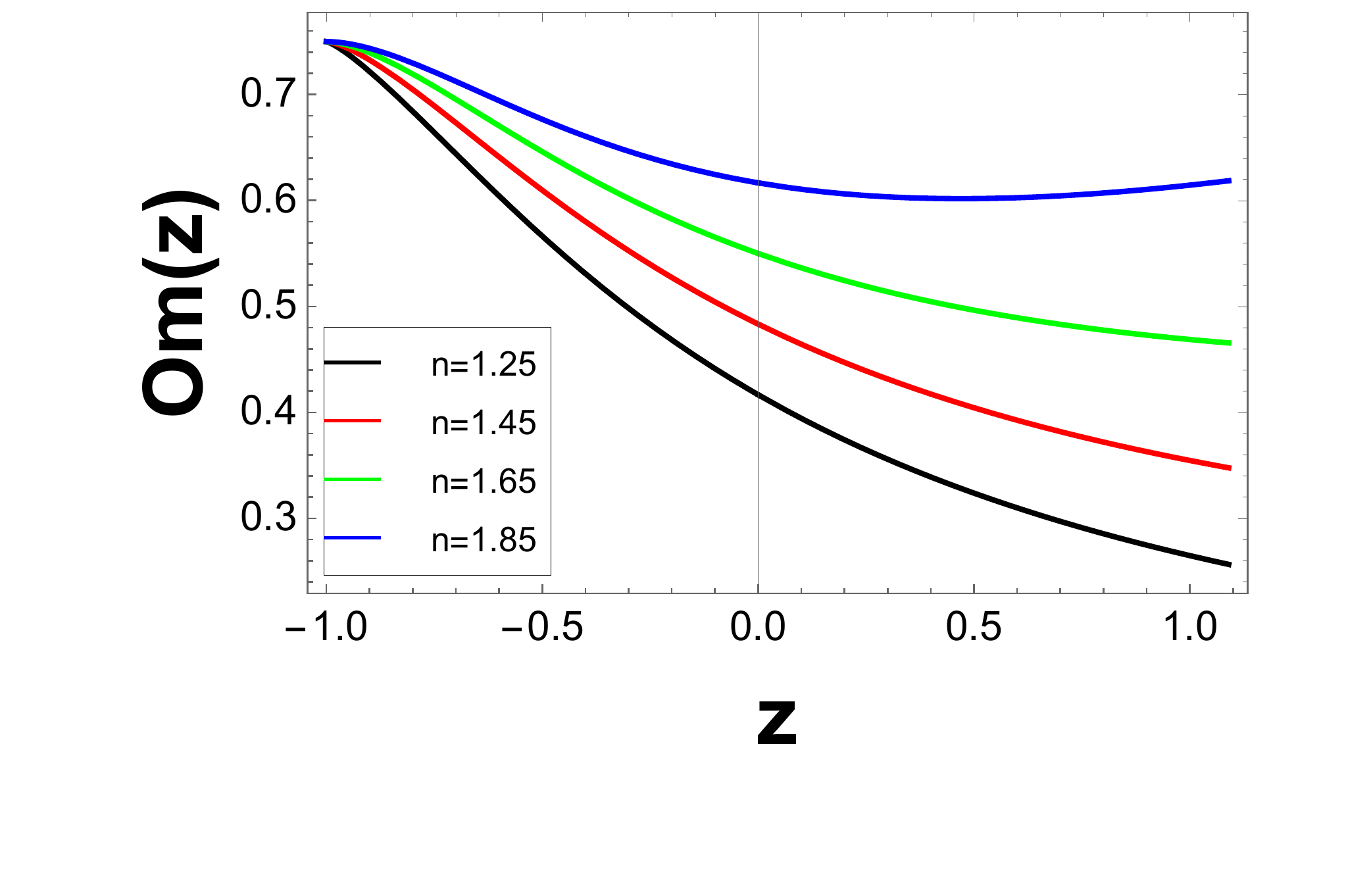} 
\end{array}
$
\end{center}
\caption{\scriptsize The plot of $Om(z)$ \textit{vs.} $z$.}
\end{figure}
The contrasting behavior of dark energy models from $ \Lambda CDM $ depend on the slope of the Om(z) diagnostic. A quintessence $(\omega >-1)$ type behavior of dark energy can be identified by its negative curvature with respect to $ z $, and a phantom type behavior $ (\omega <-1) $ can be diagnosed by its positive curvature with respect to $ z $, and a zero curvature of $Om(z)$ represents the standard $ \Lambda CDM $.\\

Figure 8 exhibits the evolution of different trajectories of the function $ Om(z) $ with respect to the redshift $ z $, corresponding to different values of model parameter $n$. From the plot of Fig. 8, we can observe that all the trajectories show negative slope, \textit{i.e.,} all the trajectories move in an upward direction as time increases or redshift decreases. The negative curvature pattern suggests that our model is behaving similar to quintessence for all values of $ n $.

\section{Observational constraints on the model parameters}

\qquad According to the current survey, as is well known, a wide variety of observational data including the Cosmic Microwave Background ($ CMB $) data (which is the relic radiations from the baby Universe), Slogan Digital Sky Survey ($ SDSS $) data (the observations of the distribution of galaxies with position redshift which essentially encode the fluctuations in the Universe, spectra of quasars), Type Ia supernovae data (usually known as standard candles and used to measure the expansion of Universe), Baryon Acoustic Oscillations ($ BAO $) data (which measures the structure in the Universe), and large-scale structure ($ LSS $) data (having provided a very strong tool to test our cosmological framework for many years \cite{mcc}) produce for various measurements of challenging issues in our Universe like the evolution of the Universe, properties of dark matter and dark energy. More significantly, the growth rate of structure tests are independent and complementary to the constraints, which may be obtained from the analysis of the temperature and polarization fluctuations in the CMB and other observations such as Type Ia supernovae and Baryon Acoustic Oscillations ($ BAO $). The consistency between these observational data sets function as one of the strongest reasons in favor of the current standard model, the $ \Lambda CDM $ model.\\

The most significant part of the parametric reconstruction is the assessment of the values of the parameters from the observational data. The two parameters, Hubble parameter $ H_0 $ and the model parameter $ n $, are involved in our model. Here, we use $ H(z) $, $ SNeIa $ and $ BAO $ data set for the statistical analysis. In the following subsection, we constrain the model parameter $ n $ with the observational $ H(z) $, $ SNeIa $ (Union 2.1 compilation), $ BAO $ data set, and the joint data set $ H(z)$ + $ SNeIa $ and $ H(z)$ + $ SNeIa $ + $ BAO $  respectively, for which the corresponding values of Hubble parameter  $ H_0 $  can also be constrained. 

\begin{figure}[tbph]
\begin{center}
$%
\begin{array}{c@{\hspace{.1in}}c}
\includegraphics[width=3.0 in, height=2.5 in]{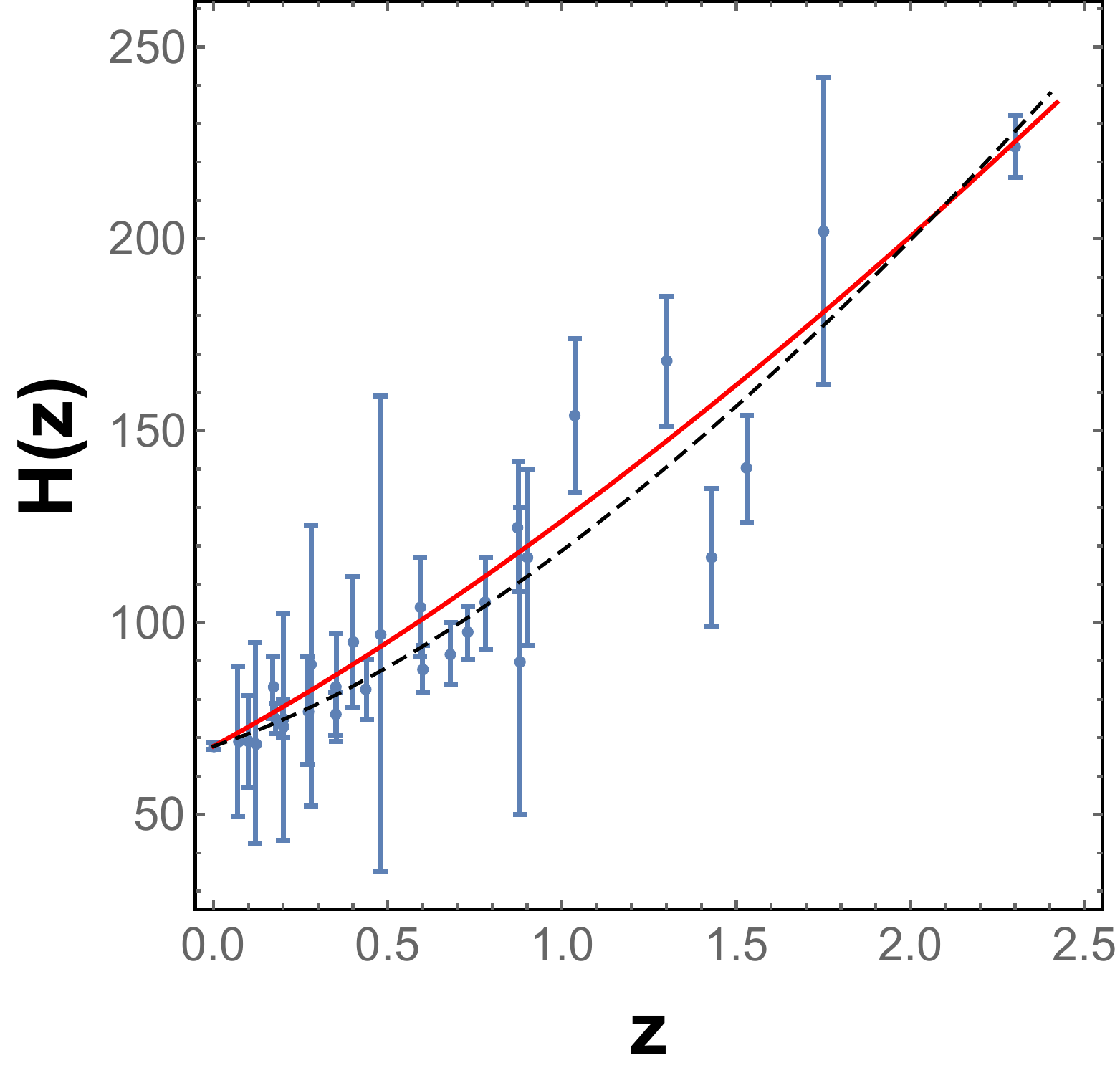} & %
\includegraphics[width=3.0 in, height=2.5 in]{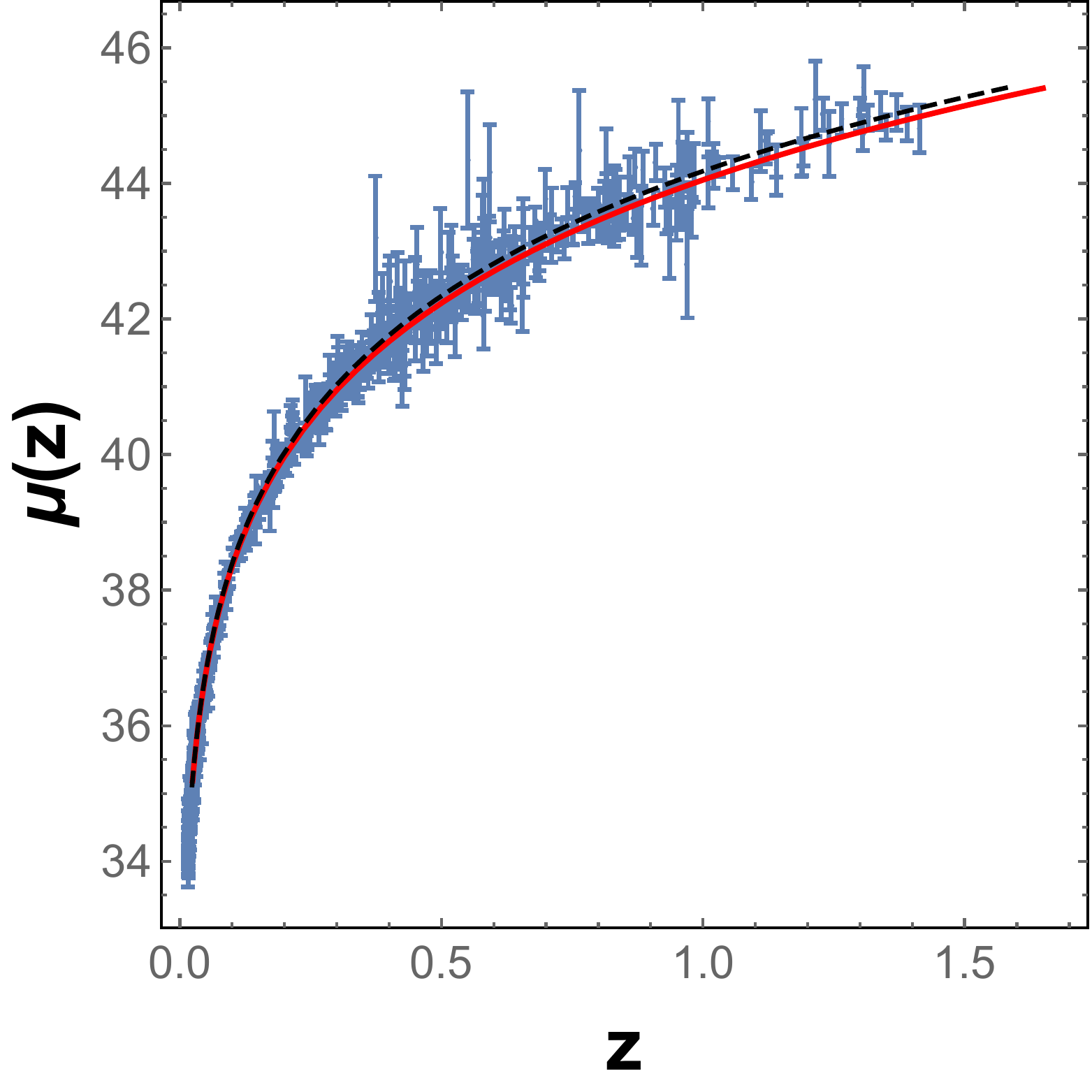} \\ 
\mbox (a) & \mbox (b)%
\end{array}%
$%
\end{center}
\caption{\scriptsize Figures a and b show the comparison of our model and the $ \Lambda CDM $ model with error bar plots of Hubble data set and $ SNeIa $ Union 2.1 compilation data set, respectively. Red lines indicate our model and dashed black lines indicate the $ \Lambda CDM $ model in both plots.}
\end{figure}

\subsection{Hubble parameter H(z)}
\qquad In this subsection, we compare our model with the $ 29 $ points of $ H(z)$ data set \cite{Hz} in the redshift range $ 0.1\leqslant z\leqslant 2.5 $ and compare with the $ \Lambda CDM $ model. We choose the value of the current Hubble constant from Planck 2014 results \cite{Hz-Planck} as $ H_{0}=67.8 $ $ Km/s/Mpc $ to complete the data set.\\

The mean value of the model parameter $ n $ determined by minimizing the corresponding chi-square value \textit{i.e.} $ \chi^2_{min} $, which is equivalent to the maximum likelihood analysis is given by

\begin{equation}\label{oc1}
\chi _{OHD}^{2}(p_{s})=\sum\limits_{i=1}^{28}\frac{[H_{th}(p_{s},z_{i})-H_{obs}(z_{i})]^{2}}{\sigma _{H(z_{i})}^{2}},
\end{equation}
where $ OHD $ is the observational Hubble data set. $ H_{th} $ and $ H_{obs}$ represent the theoretical and observed value of Hubble parameter $ H $ of our model whereas $ p_{s} $ refers to the model parameter $ n $. The standard error in the observed value is denoted by $ \sigma_{H(z_{i})} $. 
\begin{figure}[tbph]
\begin{center}
$%
\begin{array}{c@{\hspace{.1in}}c}
\includegraphics[width=3.0 in, height=2.5 in]{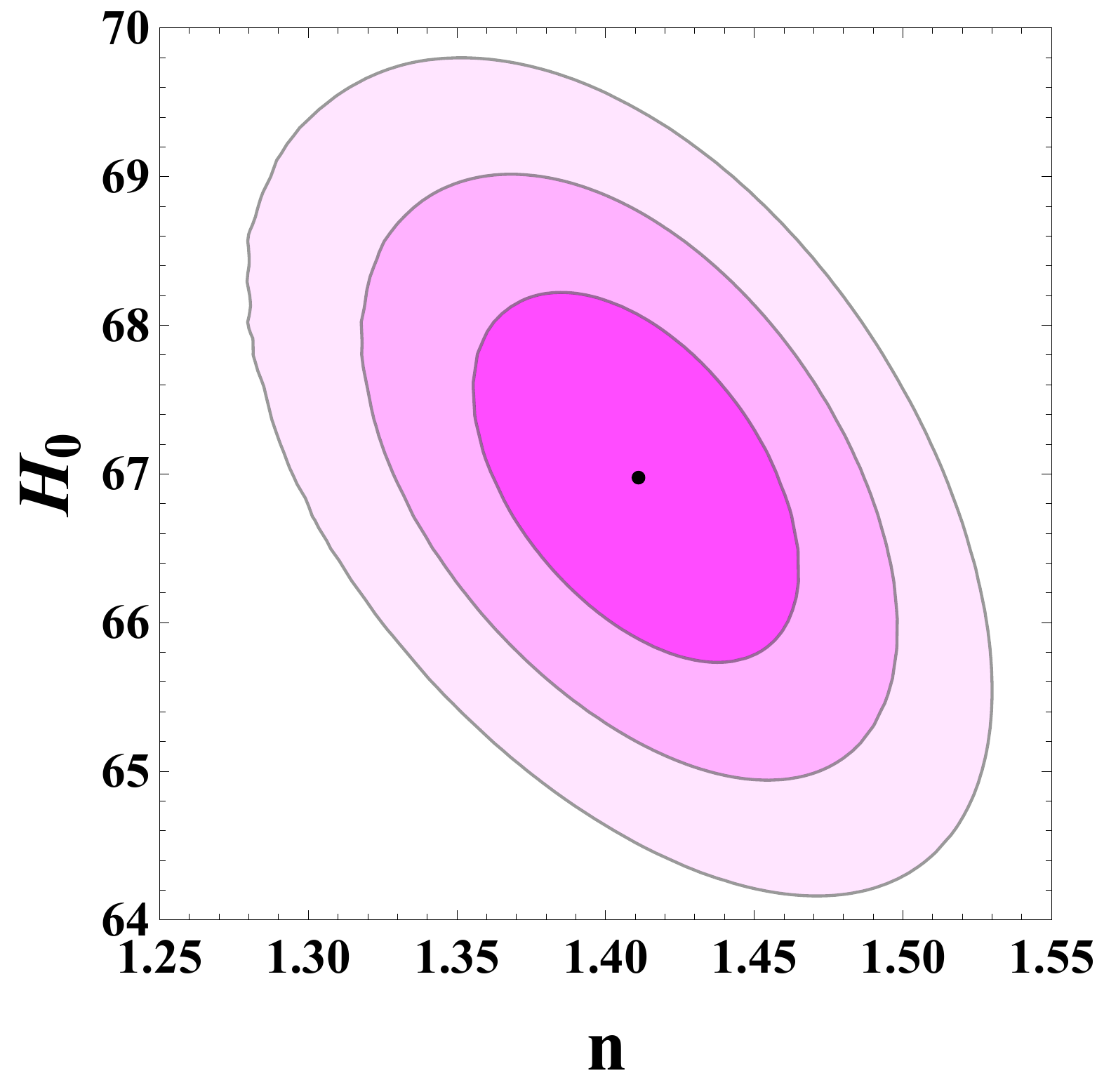}\\
\end{array}%
$%
\end{center}
\caption{\scriptsize Figure shows the likelihood contour in the $ n $-$ H_{0} $ plane for $ H(z)$ dataset. The dark shaded region shows the 1 $ \sigma $ error, light shaded region shows 2 $ \sigma $ error and ultra light shaded region shows 3 $ \sigma $ error. Black dot represents the best fit value of model parameter $ n $ and the values of $ H_{0} $ in the plot.  Here,  $ H_0 $ is in the unit of $ Km/s/Mpc$.}
\end{figure}
\subsection{Type Ia Supernova}
\qquad In this subsection, we fit our model with the $ 580 $ points of the Union 2.1 compilation $ SNeIa $ data set \cite{SNeIa} and compare with the $ \Lambda CDM $ model. We choose the value of the current Hubble constant from the Planck 2014 results \cite{Hz-Plank} as $ H_{0}=67.8$ $ Km/s/Mpc $ to complete the data set. 
\begin{equation}\label{oc2}
\chi _{OSN}^{2}(\mu_0,p_{s})=\sum\limits_{i=1}^{580}\frac{[\mu_{th}(\mu_0,p_{s},z_{i})-\mu_{obs}(z_{i})]^{2}}{\sigma _{\mu(z_{i})}^{2}},
\end{equation}
where $  OSN $ is the observational $ SNeIa $ data set. $\mu_{th}$ and $\mu_{obs}$ are the theoretical and observed distance modulus of the model. The standard error in the observed value is denoted by $\sigma_{\mu(z_{i})}$. The distance modulus  $\mu(z)$ is defined by 
\begin{equation}\label{oc3}
\mu(z)= m-M = 5Log D_l(z)+\mu_{0},
\end{equation}
where $m$ and $M$ indicate the apparent and absolute magnitudes of a standard candle, respectively. The luminosity distance $ D_l(z) $ and the nuisance parameter $\mu_0$ are defined as: 
\begin{equation}\label{oc4}
D_l(z)=(1+z)H_0\int_0^z \frac{1}{H(z^*)}dz^*,
\end{equation}
and
\begin{equation}\label{oc5}
\mu_0= 5Log\Big(\frac{H_0^{-1}}{Mpc}\Big)+25,
\end{equation}
respectively. In order to calculate the luminosity distance, we have restricted the series of $H(z)$ up to the tenth term, then integrating the approximate series to obtain the luminosity distance.\\

The left panel of Fig. 9 displays the best fitting curve of our model compared with the $ \Lambda CDM$ model for the $H(z)$ data set and the right panel shows best fitting curve of our model compared with the $ \Lambda CDM $ model for $ SNeIa $ data set. 

\subsection{Baryon acoustic oscillations}

\qquad In the early Universe, before matter decouples, baryons and photons form a plasma through which sound waves can propagate. Sound waves can leave a very distinct imprint on the statistical properties on matter that we can call $ BAO $. It measures the structures in the Universe from very large scales which allow us to understand dark energy better. In this study we adopt a sample of $ BAO $ distances measurements from different surveys, namely $ SDSS(R) $ \cite{padn}, the $ 6dF $ Galaxy survey \cite{6df}, $ BOSS\, CMASS $ \cite{boss} and three parallel measurements from $ WiggleZ $  survey  \cite{wig}.\\

In context of $ BAO $ measurements, the distance redshift ratio $d_z$ is given by
\begin{equation}\label{drr}
d_z=\frac{r_s(z_*)}{D_v(z)},
\end{equation}
where $ r_s(z_*) $ is defined as the co-moving sound horizon at the time when photons decouple, $ z_* $ indicates the photons decoupling redshift and is taken as $ z_*=1090 $ according to the Planck 2015 results \cite{adep}. Also $r_s(z_*)$ is assumed to be the same as it is considered in \cite{waga}. Further, the dilation scale is denoted by $ D_v(z) $ and is given by the relation $D_v(z)= \big(\frac{d^2_A(z) z}{H(z)}\big)^{\frac{1}{3}}$, where $d_A(z)$ is the angular diameter distance.\\

The  value of $\chi _{BAO}^{2}$ corresponding to $ BAO $ measurements is given by \cite{gio}
\begin{equation}\label{OBAO}
\chi _{BAO}^{2}= A^{T}C^{-1}A,
\end{equation}
where $ A $ is a matrix given by
 \[
   A=
  \left[ {\begin{array}{cc}
   \frac{d_A(z_*)}{D_v(0.106)}-30.84  \\
   \frac{d_A(z_*)}{D_v(0.35)}-10.33 \\
   \frac{d_A(z_*)}{D_v(0.57)}-6.72 \\
   \frac{d_A(z_*)}{D_v(0.44)}-8.41 \\
   \frac{d_A(z_*)}{D_v(0.6)}-6.66 \\
   \frac{d_A(z_*)}{D_v(0.73)}-5.43 \\
  \end{array} } \right]
\]
and $ C^{-1} $ is the inverse of covariance matrix \cite{gio} given by
\[
  C^{-1}=
  \left[ {\begin{array}{cccccc}
   0.52552 & -0.03548 & -0.07733 & -0.00167 & -0.00532 & -0.00590\\
   -0.03548 & 24.97066 & -1.25461 & -0.02704 & -0.08633 & -0.09579\\
   -0.07733 & -1.25461 & 82.92948 & -0.05895 & -0.18819 & -0.20881\\
   -0.00167 & -0.02704 & -0.05895 & 2.91150 & -2.98873 & 1.43206\\
  -0.00532 & -0.08633 & -0.18819 & -2.98873 & 15.96834 & -7.70636\\
  -0.00590 & -0.09579 & -0.20881 & 1.43206 & -7.70636 & 15.28135\\
  \end{array} } \right]
  \]
adopting the correlation coefficients presented in \cite{hing}.\\

\begin{figure}[tbph]\label{hsnjoint}
\begin{center}
$%
\begin{array}{c@{\hspace{.1in}}c}
\includegraphics[width=3.0 in, height=2.5 in]{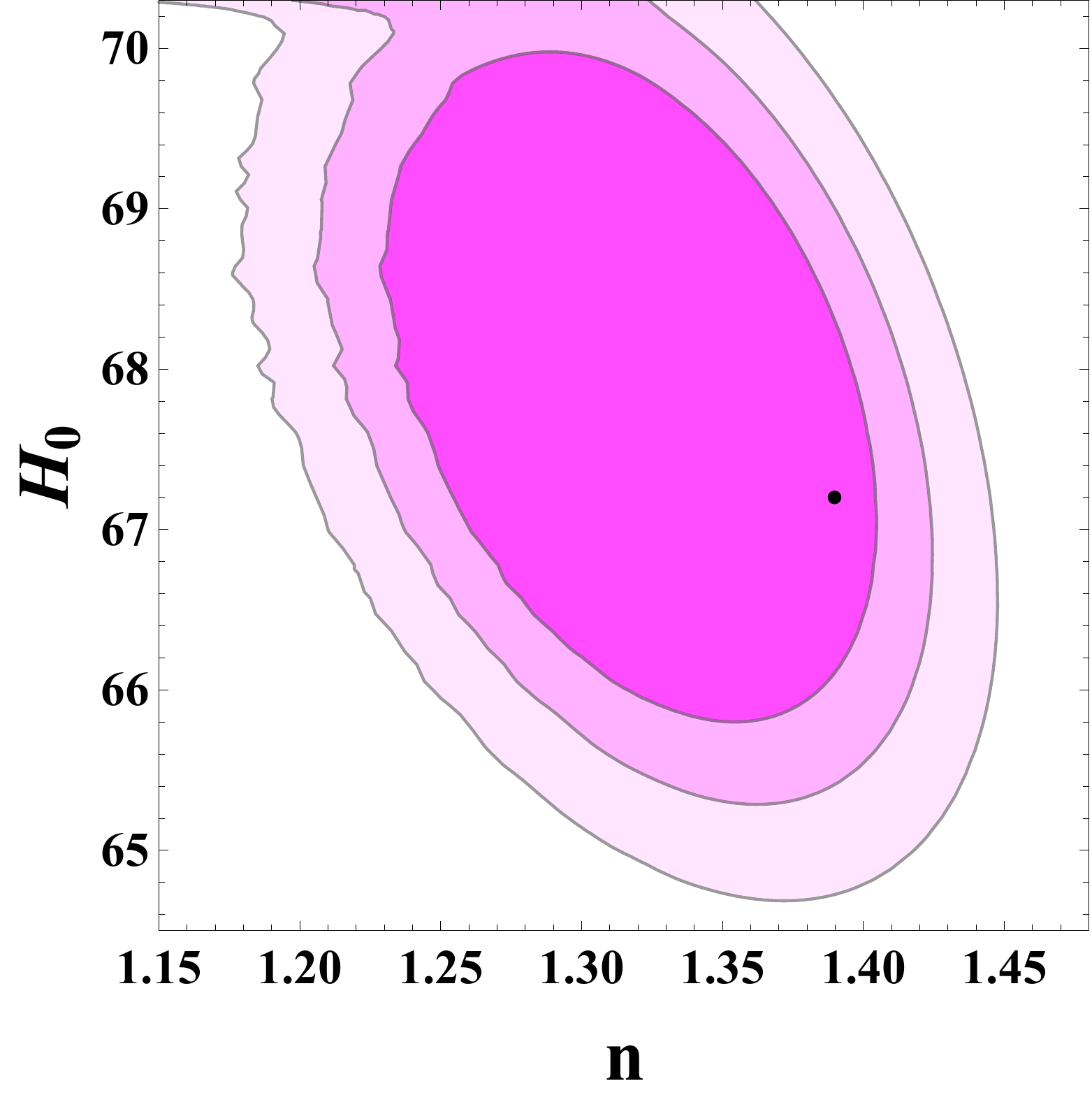} & 
\includegraphics[width=3.0 in, height=2.5 in]{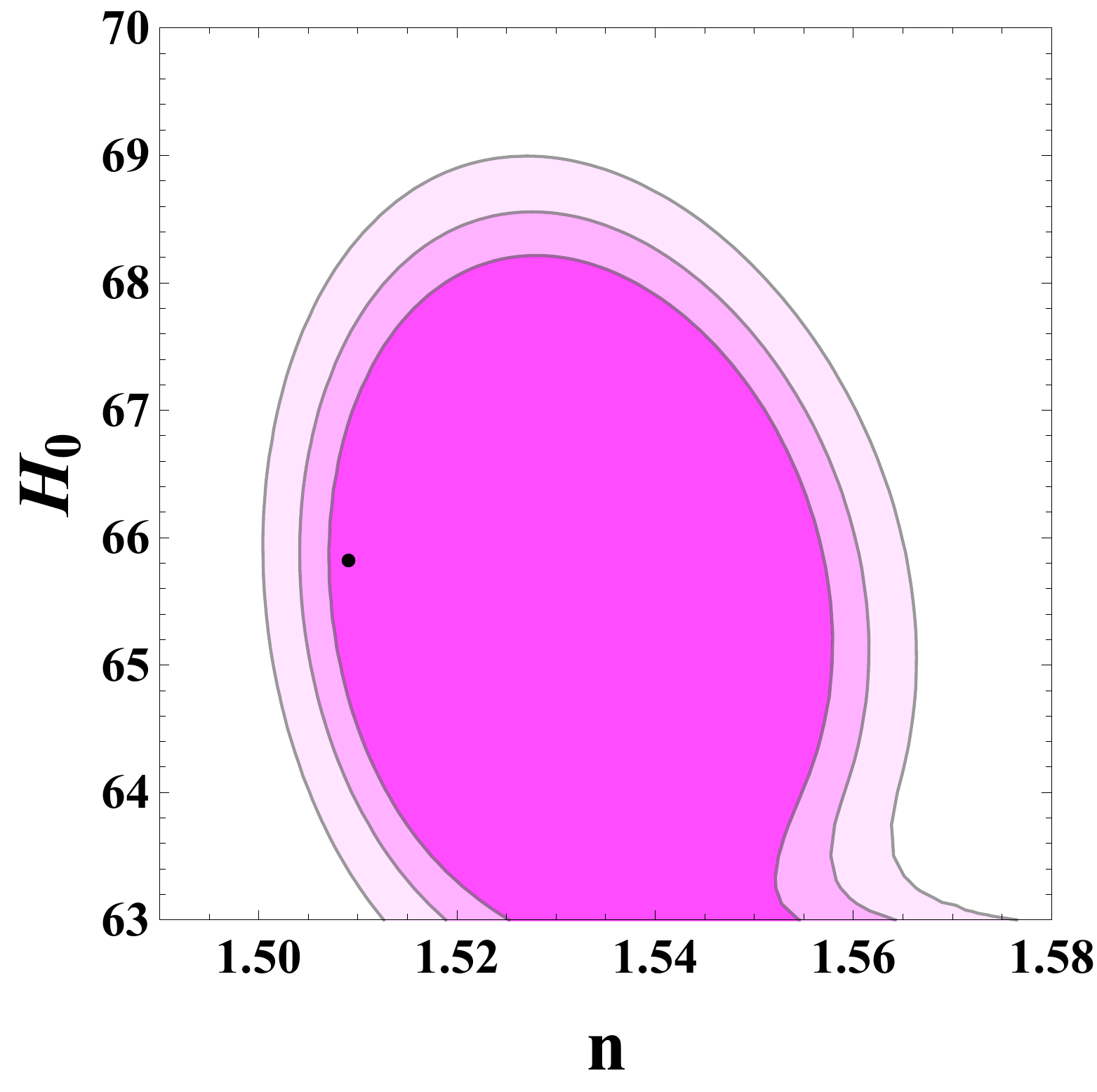} \\ 
\mbox (a) & \mbox (b)%
\end{array}%
$%
\end{center}
\caption{\scriptsize Figures (a) and (b) show the likelihood contours in the $ n $-$ H_{0} $ plane for joint analysis $ H(z)$ + $ SNeIa $  and $ H(z)$ + $ SNeIa $  + $ BAO $ respectively. The dark shaded region shows the $1 \sigma $ error, light shaded region shows the $ 2 \sigma $ error and ultra light shaded region shows the $ 3  \sigma $ error. Black dots represent the best fit values of the model parameter $ n $ and the values of $ H_{0} $ in both plots.  Here, $ H_0 $ is in the units of km/s/Mpc}
\end{figure}

The likelihood contours for the parameters $ n $ and $ H_{0} $ with $ 1 \sigma $, $ 2 \sigma $ and $ 3  \sigma $ errors in the $ n $-$ H_{0} $ plane are shown in Figs. 10 and 11. The best fit values of $ n $ are found to be $ 1.41 $, $ 1.30 $, $ 1.509 $, $ 1.39 $ and $ 1.509 $ according to the Hubble data set and $ SNeIa $  (Union 2.1 compilation data set), $ BAO $ , joint data set $ H(z)$ + $ SNeIa $  and the joint data set $ H(z)$ + $ SNeIa $  + $ BAO $   for which the corresponding best fit values of $ H_0 $ are constrained as $ 66.9762 $, $68.5583$, $68.8486$, $ 67.2050 $ and $65.8202$, respectively (see Table 2).

\begin{table}
\caption{ Summary of the numerical results for flat universe.}
\begin{center}
\label{tabparm}
\begin{tabular}{l c c c r} 
\hline\hline
\\ 
{Data} & \,\,\,\,\, $ \chi^2_{min} $ \,\,\,  &  \,\, \, Hubble parameter  $ H_0 $ {\footnotesize(km/s/Mpc)} \,\,\,   &  \,\,\, Parameter $  n $ \,\,\,\\ 
\\
\hline 
\\
{$ H(z) $ {\footnotesize{(29 points data )} }}    &    $ 24.5790 $   &   $ 66.9762 $   &   $ 1.410 $
\\
\\
{$ SNeIa $ {\footnotesize{(Union 2.1 compilation data)}}}   &    $ 586.173 $   &   $ 66.6213 $   &   $ 1.390 $ 
\\
\\
{$ BAO $}   &   $ 29.4699 $   &   $ 68.8486 $   &   $ 1.509 $ 
\\
\\
{$ H(z) $ + $ SNeIa $ }  &  $ 611.0960 $  &  $ 67.2050 $  &  $ 1.390 $
\\
\\
{$ H(z)$ + $ SNeIa $ + $ BAO $}   &   $ 673.6844 $   &   $ 65.8202 $   &   $ 1.509 $
\\
\\ 
\hline\hline  
\end{tabular}    
\end{center}
\end{table}

\section{Discussions and Conclusions}
\qquad In this article, we have presented a $ \Lambda (t) $ cosmology model obtained by a simple parametrization of the Hubble parameter in a flat FLRW space-time in $ f(R,T) $ modified gravity theory. We have studied the most simple form of $ f(R,T)$ function that can explain the non-minimal coupling between geometry and matter present in the Universe. The field equations have been derived by taking the functional form of $ f(R,T)=f(R)+f(T) $ into consideration, which leads to general relativistic field equations with a trace $ T $ dependent term. We called this term the cosmological constant $ \Lambda (T) $ in this study. To obtain the exact solution of the cosmological field equations, we have endorsed a parametrization of the Hubble parameter $ H $ that yields a time dependent deceleration parameter $ q(t) $. Comprehensive observations have been recorded for our obtained model based on the above-mentioned information.

\begin{itemize}
\item [(i)] In order to study a cosmological model capable of explaining the recent astronomical observations of accelerating expansion of the Universe with a decelerating phase of evolution in the past, we have considered a geometrical parametrization of the Hubble parameter $H$ used by Singh \cite{jps} and Banerjee et al.  \cite{ban}, which leads to a variable deceleration parameter $q$. The obtained form of $q$ describes both the scenario of early deceleration and present acceleration. The behavior of the geometrical parameters $a$, $H$ and $q$ at two extremities ($t\rightarrow 0,t\rightarrow \infty $) have been analyzed in Table 1.

\item [(ii)] For the considered parametrization of $H$, the different phases of evolution of the deceleration parameter has been examined. From the expression of $q(z)$, we have found the range of the deceleration parameter, \textit{i.e.}, $ q\in \lbrack n-1,-1] $, which clearly shows the signature flipping behavior because the model parameter $n>1$. For a close view on $ q $, we can observe the decelerating to accelerating regimes of the Universe depending on the variation of the model parameter $n$ in Fig. 1. As the values of $n$ increase from $1.25$ to $1.85$, the phase transition redshift $ z_{tr} $ could be delayed.

\item [(iii)] To discuss the role of the $ f(R,T) $ coupling constant $ \lambda $ played in the evolution of the EoS parameter $\omega $, we fix the value of $ n = 1.45 $ and vary $ \lambda $. In Fig. 2, we have examined the special characteristic of the coupling constant $ \lambda $, and observed the variation in $\omega$ as $\lambda $ takes both negative and positive values. This shows the contribution of $ f(R,T) $ gravity in this model on considering the acceptable range of $ \lambda $.

\item [(iv)] In Table I, we have shown the behavior of  $ a$, $ H $, $ q $, $ \rho $, $ p $, $ \omega $, and $ \Lambda $ at $ t\rightarrow 0 $ and $ t\rightarrow \infty $  and studied the physical significance of $ \rho $, $ p $, $ \omega $, and $ \Lambda $ with respect to the redshift $ z $ in Sect. 3.3. In Fig. 3, the energy density and isotropic pressure reduces from their dense state to a constant value, which depends on $ \lambda $. Figure 3b depicts the isotropic pressure $ p $ starting from a very large value at the initial singularity and approaching to $ \frac{-3\alpha^{2}}{(A+1)} $ in the future $ z\rightarrow -1 $ for some specific values of $ n $. The negative values of cosmic pressure are corresponding to the cosmic acceleration according to standard cosmology. Hence our model exhibits accelerated expansion at present as well as in infinite future. The model is consistent with the structure formation of the Universe. In Fig. 4a, for all values of $ n $ and a fixed value $ \lambda=-130 $, the EoS parameter $ \omega $ transits from positive to negative and ultimately approaches the quintessence region, which suggest that matter in the Universe behaves like perfect fluid initially and as dark energy in late time. In Fig. 4b, the cosmological constant $ \Lambda $ starts decreasing from a very high value at high redshift and approaches  a small positive value at present epoch $ (z\rightarrow 0) $, which is in good agreement with the current observations \cite{per,rie,ton,clo}.

\item [(v)] Next, we have compared our dark energy model with standard $ \Lambda CDM $ model by examining the behavior of the other geometrical parameters, \textit{e.g.}, the jerk $ j $, snap $ s $ and lerk $ l $ parameters. From Fig. 5a, it can be seen that, for every value of $ n $, our model behaves different from the $ \Lambda CDM $ model at the present time $ z=0 $, but in the late future $ j\rightarrow 1 $, which is in accordance with the $ \Lambda CDM $ model. In addition to $ j $, the behaviors of the snap $ s $ and lerk $ l $ parameters are graphically demonstrated in Fig. 5. The snap parameter $ s $ shows one transition from negative to positive throughout its evolution with respect to the redshift $ z $, while the lerk parameter $ l $ is decaying in nature with no transition.

\item [(vi)] In Sect. 4, some physical analysis and geometrical diagnostics of the model has been studied. The physical viability of the model has been analyzed by verifying the energy conditions of our model. In Fig. 6, it can be seen easily that NEC and DEC hold good but SEC fails for all values of the model parameter $ n $ and the fixed value $ \lambda= -130 $.

\item [(vii)] In Sect. 4.2, the Fig. 7 represents the time evolution of four trajectories for different values of $ n $ in $ \{s,r\} $ and $ \{q,r\} $ plane diagram. The directions of $ s-r $ trajectories in the plane diagram are represented by arrows, showing different dark energy models and they ultimately approach $ \Lambda CDM $ (see Fig. 7a). In the $ q-r $ plane diagram, the evolution of the trajectories for different values of $ n $, commences in the vicinity of $ SCDM $, and as time evolves, the trajectories of $ q-r $ approach the steady state model $ SS $ (see Fig. 7b).

\item [(viii)] Also, one more geometrical diagnostic has been interpreted to gain understanding of the different dark energy models for every value of $ n $. A plot of the $ Om(z)$ against redshift $z$ has been displayed in Fig. 8. All the
trajectories of $ Om(z) $ exhibit a negative slope, which suggests that our model is behaving similar to a quintessence model for all $ n $ and in the late time, \textit{i.e.,} $ z\rightarrow -1 $, $ Om(z)\rightarrow k $, where $ k $ is a positive finite quantity. This means that our model may correspond to $ \Lambda CDM $ in the future.

\item [(ix)] The likelihood contours for the model parameters $ n $ and $ H_{0} $ with $ 1 \sigma $, $ 2 \sigma $ and $ 3  \sigma $ errors in the $ n $-$ H_{0} $ plane are shown in Figs. 10 and 11. The model parameter $ n $ is constrained using the $ 29 $ points of the $ H(z) $ and Union 2.1 compilation data. The obtained model is in good agreement with the $ H(z) $ and $ SNeIa $ (Union 2.1 compilation data) and nearly follows the $ \Lambda CDM $ behavior (see Fig. 9). The constrained best fit values of the model parameters $ n $ are $ 1.41 $, $ 1.30 $, $ 1.509 $, $ 1.39 $ and $ 1.509 $ according to the Hubble data $ H(z)$ and $ SNeIa $  (Union 2.1 compilation data), $ BAO $, joint data $ H(z)$ + $ SNeIa $  and $ H(z) $ + $ SNeIa $  + $ BAO $ for which the corresponding best fit values of $ H_0 $ are evaluated to be $ 66.9762 $, $ 68.5583 $, $ 68.8486 $, $ 67.2050 $ and $ 65.8202 $ respectively (see Table 2).

\end{itemize}
With the above points, we conclude that our $ \Lambda $-cosmological model in $f(R,T)$ gravity within the framework of the FLRW metric is different from other $ \Lambda $-cosmological models in $ f(R,T) $ gravity discussed by other researchers mentioned in the introduction. Therefore, our research work may be fruitful for further investigation.
\vskip0.2in
\textbf{\noindent Acknowledgements } The authors express their gratitude to Prof. M. Sami and Prof. S. G. Ghosh, Centre for Theoretical Physics, Jamia Millia Islamia, New Delhi, India, and Prof. J. P. Saini, Director, NSIT, New
Delhi, India, for some fruitful discussions, and providing necessary facilities to complete the work. The author JKS expresses his gratitude to the Department of Mathematical Sciences, University of Zululand, Kwa-Dlangezwa 3886, South Africa, and Department of Mathematics, Statistics and Computer Sciences, University of KwaZulu-Natal, Westville 4001, South Africa, for some fruitful discussions with Prof. S. D. Maharaj, their financial support and providing necessary facilities as
well as hospitality; part of the work was completed there. Moreover, the work of KB was supported by the JSPS KAKENHI Grant Number JP25800136 and Competitive Research Funds for Fukushima University Faculty (17RI017 and 18RI009). The authors also express their gratitude to the referee for valuable comments and suggestions.

\end{document}